\documentclass[a4paper,fleqn,usenatbib]{mnras}

\usepackage[T1]{fontenc}
\usepackage{ae,aecompl}

\usepackage{graphicx}	
\usepackage{amsmath}	
\usepackage{amssymb}	

\title[Small-scale structure in MRI-driven turbulence]{Universal small-scale structure in turbulence driven by magnetorotational instability}

\author[V. Zhdankin et al.]{
Vladimir Zhdankin,$^{1}$\thanks{E-mail: zhdankin@jila.colorado.edu}
Justin Walker,$^{2}$
Stanislav Boldyrev,$^{2,3}$
Geoffroy Lesur$^{1,4,5}$\thanks{Visiting fellow}
\\
$^{1}$JILA, NIST and University of Colorado, 440 UCB, Boulder, Colorado 80309, USA\\
$^{2}$Department of Physics, University of Wisconsin-Madison, 1150 University Avenue, Madison, Wisconsin 53706, USA\\
$^{3}$Space Science Institute, Boulder, Colorado 80301, USA\\
$^{4}$Univ. Grenoble Alpes, IPAG, F-38000 Grenoble, France\\
$^{5}$CNRS, IPAG, F-38000 Grenoble, France\\
}

\date{Accepted XXX. Received YYY; in original form ZZZ}

\pubyear{2016}

\begin{document}
\label{firstpage}
\pagerange{\pageref{firstpage}--\pageref{lastpage}}
\maketitle

\begin{abstract}
The intermittent small-scale structure of turbulence governs energy dissipation in many astrophysical plasmas and is often believed to have universal properties for sufficiently large systems. In this work, we argue that small-scale turbulence in accretion disks is universal in the sense that it is insensitive to the magnetorotational instability (MRI) and background shear, and therefore indistinguishable from standard homogeneous magnetohydrodynamic (MHD) turbulence at small scales. We investigate the intermittency of current density, vorticity, and energy dissipation in numerical simulations of incompressible MHD turbulence driven by the MRI in a shearing box. We find that the simulations exhibit a similar degree of intermittency as in standard MHD turbulence. We perform a statistical analysis of intermittent dissipative structures and find that energy dissipation is concentrated in thin sheet-like structures that span a wide range of scales up to the box size. We show that these structures exhibit strikingly similar statistical properties to those in standard MHD turbulence. Additionally, the structures are oriented in the toroidal direction with a characteristic tilt of approximately 17.5 degrees, implying an effective guide field in that direction. \\
\end{abstract}

\begin{keywords}
turbulence -- plasmas -- MHD -- accretion discs -- instabilities
\end{keywords}



\section{Introduction}

Turbulence plays a central role in accretion disks by governing the angular momentum transport responsible for mass accretion onto compact objects \citep{balbus_hawley_1998}, amplifying magnetic fields via the dynamo \citep{brandenburg_etal_1995}, and intermittently dissipating magnetic energy, which may lead to non-thermal particle acceleration and X-ray flares \citep{baganoff_etal_2001, markoff_etal_2001, mcclintock_remillard_2006}. Many questions on the nature of turbulence in accretion disks remain, and it is important to understand to what extent insights from standard magnetohydrodynamic (MHD) turbulence can be adapted. In this work, we use the term standard MHD turbulence to refer to externally driven or decaying MHD turbulence in a periodic box in the absence of background gradients (and therefore statistically homogeneous). Standard MHD turbulence serves as a conventional testing ground for controlled numerical studies of turbulence in the inertial range.

Turbulence can be driven in ionized accretion disks by the magnetorotational instability (MRI) when a weak background magnetic field and Keplerian shear are present \citep{balbus_hawley_1991}. The dynamics are often studied in the shearing box approximation, which provides a simple framework for describing turbulence in a small co-rotating patch of the accretion disk. Previous numerical studies of MRI-driven turbulence in a shearing box have primarily focused on large-scale quantities such as the angular momentum transport coefficients, magnetic and kinetic energy spectra \citep{fromang_2010, lesur_longaretti_2011}, anisotropy \citep{nauman_blackman_2014, murphy_pessah_2015}, and dynamo cycles \citep{bai_stone_2013}. Physical and numerical degrees of freedom such as the net flux \citep[e.g.,][]{salvesen_etal_2016}, stratification \citep[e.g.,][]{simon_etal_2011}, dissipation coefficients \citep{fleming_etal_2000, fromang_etal_2007, lesur_longaretti_2007}, box size \citep{bodo_etal_2008, bodo_etal_2011, simon_etal_2012, shi_etal_2016}, and boundary conditions were also explored. These numerical studies uncovered many important features of turbulence in a shearing box and reproduce aspects of global accretion disk simulations \citep{sorathia_etal_2012}. Despite their relative physical simplicity, shearing box simulations of MRI-driven turbulence have only recently identified a putative inertial range \citep{walker_etal_2016,kunz2016}, with previous numerical simulations showing little quantitative resemblence to standard MHD turbulence \citep[e.g.,][]{fromang_2010, lesur_longaretti_2011}.

One aspect of MRI-driven turbulence that was largely neglected to date is the small-scale structure of the turbulence, involving quantities such as the energy dissipation rate, current density, and vorticity. Small-scale turbulence is often believed to be universal in the sense that it is insensitive to the driving mechanism and boundary conditions, although it may depend on the system size due to intermittency (scale-dependent inhomogeneity) and scale-dependent anisotropy \citep{goldreich_sridhar_1995, boldyrev2006, mason2012, perez2012}. Intermittency causes a large fraction of energy dissipation in MHD turbulence to be concentrated in coherent structures, such as current sheets and vorticity sheets, that occupy a small fraction of the volume \citep{zhdankin_etal_2016b, wan_etal_2016}. These structures may determine the kinetic mechanisms of energy dissipation \citep{matthaeus_etal_2015}, contribute to non-thermal particle acceleration, appear as intense observable flares \citep{zhdankin_etal_2015a, zhdankin_etal_2015b}, and cause temperature inhomogeneity that leads to mineral formation \citep{mcnally_etal_2014}.

A central open question is whether the character of MRI-driven turbulence transitions to standard MHD turbulence at sufficiently small scales (which requires large simulations to properly discern), or whether the presence of the MRI across many scales modifies the dynamics at all  scales. It is likewise unknown whether the MRI modifies scale-dependent effects such as intermittency and anisotropy; even the large-scale variability of the MRI may non-trivially reduce or increase the degree of inhomogeneity and anisotropy at smaller scales. Furthermore, whereas numerical simulations established that standard MHD turbulence is characterized by thin intermittent structures that span large scales \citep[e.g.,][]{zhdankin_etal_2013, zhdankin_etal_2014}, it is unclear whether such structures can persist in the background Keplerian shear of an accretion disk.

In this work, we take a step toward answering these questions by investigating the intermittency of small-scale quantities in numerical simulations of incompressible MHD turbulence driven by the MRI in an unstratified shearing box. We demonstrate that the system is classically intermittent by showing that the distribution of coarse-grained energy dissipation rates broadens with decreasing scale, at a rate comparable to that in hydrodynamic and MHD turbulence. We then perform a statistical analysis of intermittent dissipative structures and characterize their energetics, morphology, and orientation. We show that the statistical properties of these structures are in strong quantitative agreement with similar measurements in standard MHD turbulence driven at large scales \citep{zhdankin_etal_2014, zhdankin_etal_2016, zhdankin_etal_2016b}. These results support the picture that the small-scale dynamics of turbulence, involving the current density, vorticity, energy dissipation rate, and possibly the asymptotic energy spectrum, are insensitive to the large-scale shear and MRI for sufficiently large systems.

\section{Methodology}

We consider a periodic shearing box with $x$ representing the radial direction, $y$ representing the toroidal direction, and $z$ representing the vertical direction. The incompressible MHD equations in a shearing box read
\begin{align}
(\partial_t - q \Omega_0 x \partial_y + \boldsymbol{v} \cdot \nabla) \boldsymbol{v} - \boldsymbol{B} \cdot \nabla \boldsymbol{B} &= - \nabla P + \nu \nabla^2 \boldsymbol{v} \nonumber \\ &- 2 \boldsymbol{\Omega}_0 \times \boldsymbol{v} + q \Omega_0 v_x \hat{\boldsymbol{y}} \, , \nonumber \\
(\partial_t - q \Omega_0 x \partial_y + \boldsymbol{v} \cdot \nabla) \boldsymbol{B} - \boldsymbol{B} \cdot \nabla \boldsymbol{v} &=  \eta \nabla^2 \boldsymbol{B} - q \Omega_0 B_x \hat{\boldsymbol{y}} \, , \nonumber \\
\nabla \cdot \boldsymbol{v} &= \nabla \cdot \boldsymbol{B} = 0 \, , \label{eq:mhd1}
\end{align}
where $\boldsymbol{v}$ is the velocity field, $\boldsymbol{B} = \boldsymbol{B}_0 + \boldsymbol{b}$ is the magnetic field (where $\boldsymbol{B}_0$ is the background field and $\boldsymbol{b}$ is the fluctuating part), $\boldsymbol{\Omega}_0 = \Omega_0 \hat{\boldsymbol{z}}$ is the orbital frequency, $q = 3/2$ is the Keplerian shear parameter; we absorbed a factor of $1/\sqrt{4 \pi \rho}$ in the magnetic field, where $\rho$ is the plasma density. We focus our primary analysis on the current density $\boldsymbol{j} = \nabla \times \boldsymbol{B}$ and vorticity $\boldsymbol{\omega} = \nabla \times \boldsymbol{v}$. The local resistive and viscous energy dissipation rates per unit volume are given by $\epsilon^\eta = \eta j^2$ and $\epsilon^\nu = 2 \nu \sigma_{ij} \sigma_{ij}$, respectively, where $\eta$ is the resistivity, $\nu$ is the viscosity, and $\sigma_{ij} = ( \partial v_j/\partial x_i + \partial v_i/\partial x_j ) / 2$ is the rate-of-strain tensor. We note that terms proportional to $\Omega_0$ are lower order in gradients than the other terms, and hence can be anticipated to become unimportant at small scales, leaving only the usual MHD terms.

We consider the suite of numerical simulations described in \cite{walker_etal_2016}. The dimensions of the box are chosen to be $(L_x,L_y,L_z) = (2,4,1)$ and the simulation lattice is $1024\times1024\times512$ cells. We fix the Reynolds number $Re \equiv \Omega_0 L_z^2 / \nu = 45000$ (which differs from the conventional turbulence Reynolds number, $Re_{\rm turb} \equiv v_\text{rms} (L/2\pi)/\nu \sim 4000$) and magnetic Prandtl number $Pm \equiv \nu/\eta = 1$, while varying the superimposed vertical background field $\boldsymbol{B}_0 = B_0 \hat{\boldsymbol{z}}$, taking $B_0 \in \{ 0.03, 0.01, 0.005 \}$ (in numerical units), which correspond to initial fastest growing linear MRI modes at length scales $L_{\rm MRI} \equiv 2 \pi/k_{\rm MRI} = 2 \pi (4/\sqrt{15}) B_0/\Omega_0 \in \{ 0.195, 0.065, 0.032 \} L_z$, where $k_{\rm MRI}$ is the wavevector for the fastest growing mode \citep{balbus_hawley_1998}. Hence, the simulations have varying dimensionless parameters $\beta = \Omega_0^2 L_z^2/B_0^2$ and Els\"{a}sser number $\Lambda_\eta = B_0^2/\Omega_0 \eta$. For each simulation, we consider a minimum of 7 snapshots, each separated in time by $4 \Omega_0^{-1}$, during statistical steady state. We show results from $B_0 = 0.03$ unless otherwise stated; the results that we describe are very similar for all three simulations.

\begin{figure}
\includegraphics[width=1\columnwidth]{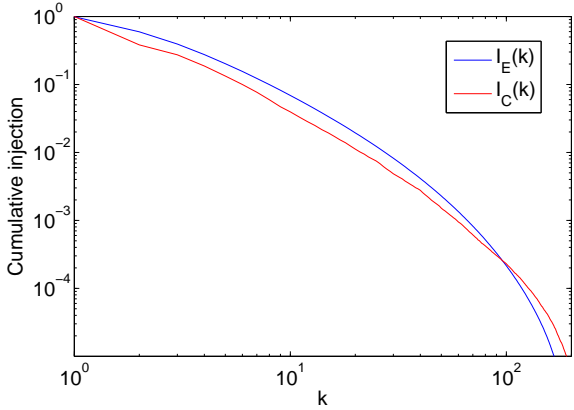}
 \centering
  \caption{\label{fig:injection} The cumulative injection rate of energy $I_E(k)$ (blue) and of cross helicity $I_C(k)$ (red) at wavenumbers greater than $k$.}
 \end{figure}
 
For reference, we now explicitly show that energy and cross-helicity injection occurs mainly at large scales. The total energy injection rate is given by $\alpha_E = q \Omega_0 \int \left(v_x v_y - B_x B_y \right)d^3 x$ while the total cross-helicity injection rate is given by $\alpha_C = (q - 2) \Omega_0 \int \left( v_x B_y - B_x v_y \right)d^3x$. The integrand can be evaluated in Fourier space to determine the injection rates at different wavenumbers, which we denote $\tilde{\alpha}_E(k)$ and $\tilde{\alpha}_C(k)$. The (normalized) cumulative injection rate for energy is then given by $I_{E}(k) = \int_k^\infty \langle \tilde{\alpha}_{E}(k) \rangle dk/\int_1^\infty \langle \tilde{\alpha}_{E}(k) \rangle dk$, where $\langle \dots \rangle$ denote ensemble average (i.e., time average). Similarly, the cumulative injection rate for cross-helicity can be characterized by $I_{C}(k) = \int_k^\infty \langle \vert \tilde{\alpha}_{C}(k) \vert \rangle dk/\int_1^\infty \langle \vert \tilde{\alpha}_{C}(k) \vert \rangle dk$, where the absolute values are taken since cross-helicity is not sign definite and is zero on average. The injection rates $I_E$ and $I_C$ are shown in Fig.~\ref{fig:injection}. We find that the injection rates strongly decrease with increasing wavenumber, becoming negligibly small compared to the cascade rate (which is unity in Fig.~\ref{fig:injection} across inertial-range scales).

\section{Results}

We first consider the distribution of coarse-grained energy dissipation rates, which is a conventional method for characterizing the scale-dependent nature of intermittency \citep{frisch1995, biskamp_1995, merrifield_etal_2005}. The coarse-grained resistive (viscous) energy dissipation rates at the $n$th level, denoted $\epsilon^\eta_n$ ($\epsilon^\nu_n$), are obtained by subdividing the simulation box $n$ times (in each direction) to obtain a set of subdomains of size $(L_x,L_y,L_z) \cdot 2^{-n}$ and measuring the average $\epsilon^\eta$ ($\epsilon^\nu$) in each of these subvolumes. The (total) coarse-grained energy dissipation rate in each subdomain is given by $\epsilon_n = \epsilon^\eta_n + \epsilon^\nu_n$.

\begin{figure*}
\includegraphics[width=1.15\columnwidth]{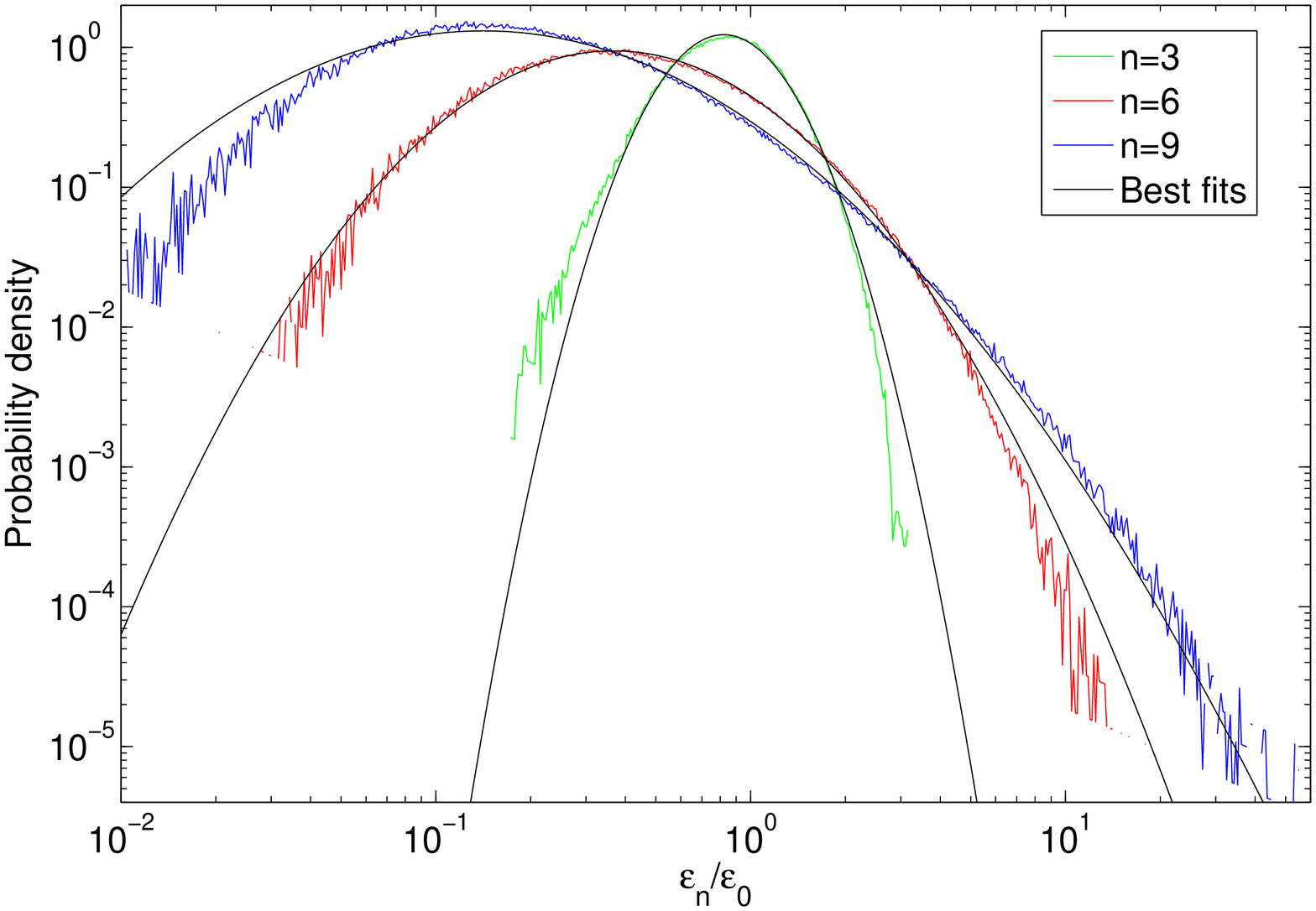}
\includegraphics[width=0.85\columnwidth]{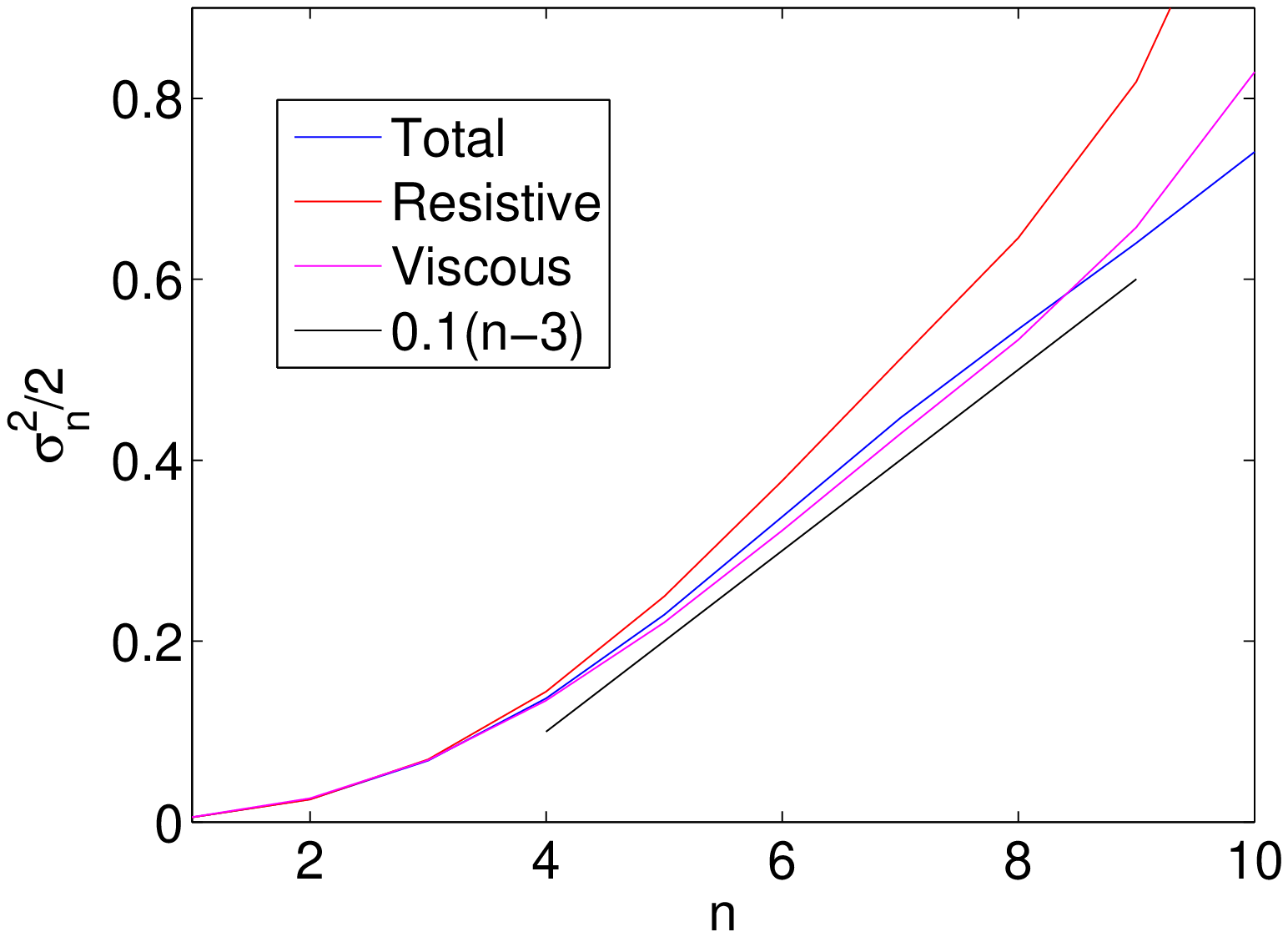}
 \centering
  \caption{\label{fig:coarsegrained} Left panel: Probability density functions for coarse-grained energy dissipation rate $\epsilon_n/\epsilon_0$ for $n \in \{3,6,9\}$ (in green, red, and blue, respectively). Best-fit log-normals are shown in black. Right panel: Scaling of $\sigma^2_n/2$ for best-fit log-normal distributions versus level $n$. Results for total dissipation are shown in blue, resistive dissipation in red, and viscous dissipation in magenta.}
 \end{figure*}

As shown in the first panel of Fig.~\ref{fig:coarsegrained}, the probability density function for coarse-grained energy dissipation rates normalized to the mean, $P(\epsilon_n/\epsilon_0)$, broadens with the number of subdivisions $n$, indicating that strong dissipative events are increasingly localized in space at smaller scales. We find that $P(\epsilon_n/\epsilon_0)$, as well as $P(\epsilon^\eta_n/\epsilon^\eta_0)$ and $P(\epsilon^\nu_n/\epsilon^\nu_0)$, are reasonably well fit for all levels $n$ by the log-normal distribution,
\begin{eqnarray}
P(\epsilon_n/\epsilon_0) = \frac{1}{\sqrt{2\pi \sigma_n^2}} \frac{\epsilon_0}{\epsilon_n} \exp{\left[-\frac{1}{2\sigma_n^2}\left(\log{\frac{\epsilon_n}{\epsilon_0}} - \mu_n \right)^2 \right]},\quad
\end{eqnarray}
where $\mu_n$ and $\sigma_n$ are the location parameter and scale parameter, respectively, with $\mu_n = -\sigma_n^2/2$ to ensure a mean of unity. As shown in the second panel of Fig.~\ref{fig:coarsegrained}, $\sigma_n^2/2 \propto n$ for a range of intermediate scales (specifically, $4 \lesssim n \lesssim 9$), consistent with the log-normal model of intermittency developed for hydrodynamic turbulence \citep{kolmogorov_1962}. The coefficient of proportionality is measured to be near $0.1$, which implies a nonzero intermittency parameter $(2/\ln{2}) 0.1 \approx 0.29$ \citep[][]{frisch1995, biskamp2003}, which is comparable to that measured in hydrodynamic turbulence \citep{sreenivasan_1993} and standard MHD turbulence \citep{zhdankin_etal_2016}.

\begin{figure*}
\includegraphics[width=1.15\columnwidth]{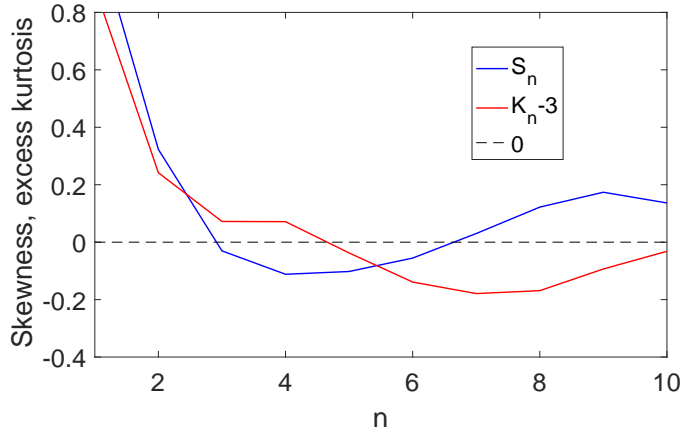}
 \centering
  \caption{\label{fig:skewness_kurtosis} Skewness $S_n$ (blue) and excess kurtosis $K_n - 3$ (red) for the logarithm of coarse-grained energy dissipation rate, $\xi_n = \log{(\epsilon_n/\epsilon_0)}$, versus level $n$.}
 \end{figure*}

The robustness of the log-normal fits can be characterized by measuring the skewness $S_n$ and kurtosis $K_n$ of the variable $\xi_n = \log{(\epsilon_n/\epsilon_0)}$, defined by
\begin{align}
S_n &= \frac{\langle (\xi_n - \bar{\xi}_n)^3 \rangle}{\langle (\xi_n - \bar{\xi}_n)^2 \rangle^{3/2}} \nonumber \\
K_n &= \frac{\langle (\xi_n - \bar{\xi}_n)^4 \rangle}{\langle (\xi_n - \bar{\xi}_n)^2 \rangle^{2}}
\end{align}
where $\bar{\xi}_n = \langle \xi_n \rangle$ and angled brackets indicate an ensemble average. For a normal distribution in $\xi_n$, which corresponds to a log-normal distribution in $\epsilon_n/\epsilon_0$, we expect that $S_n = K_n - 3 = 0$. We show the skewness $S_n$ and the excess kurtosis $K_n - 3$ in Fig.~\ref{fig:skewness_kurtosis}; since both quantities are significantly less than unity, the log-normal distribution is indeed a good fit to the overall data for energy dissipation rates. We note that the skewness increases from negative to positive in the inertial range, while the excess kurtosis decreases from positive to negative. We find that the quantities averaged for $n \ge 3$ are $\langle S_n \rangle_{n \ge 3} \approx 0.02$ and $\langle K_n \rangle_{n \ge 3} - 3 \approx -0.06$. Deviations from the log-normal distribution could be attributed to a number of factors, including corrections from a relatively short inertial range and alternative models of intermittency in MHD turbulence (e.g., the log-Poisson models considered in \citep{chandran_etal_2015, mallet_etal_2016}).

\begin{figure}
\includegraphics[width=\columnwidth]{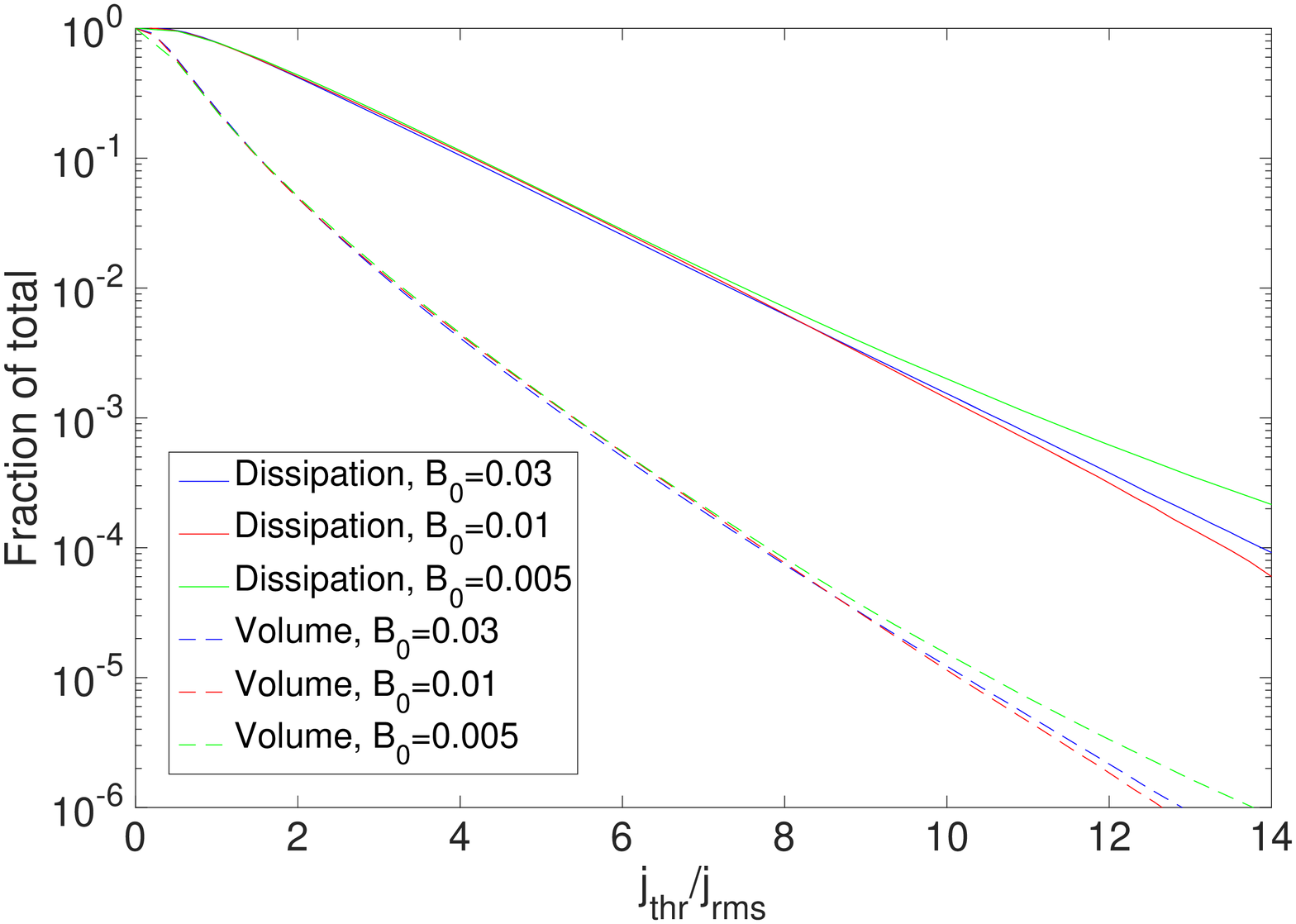}
\includegraphics[width=\columnwidth]{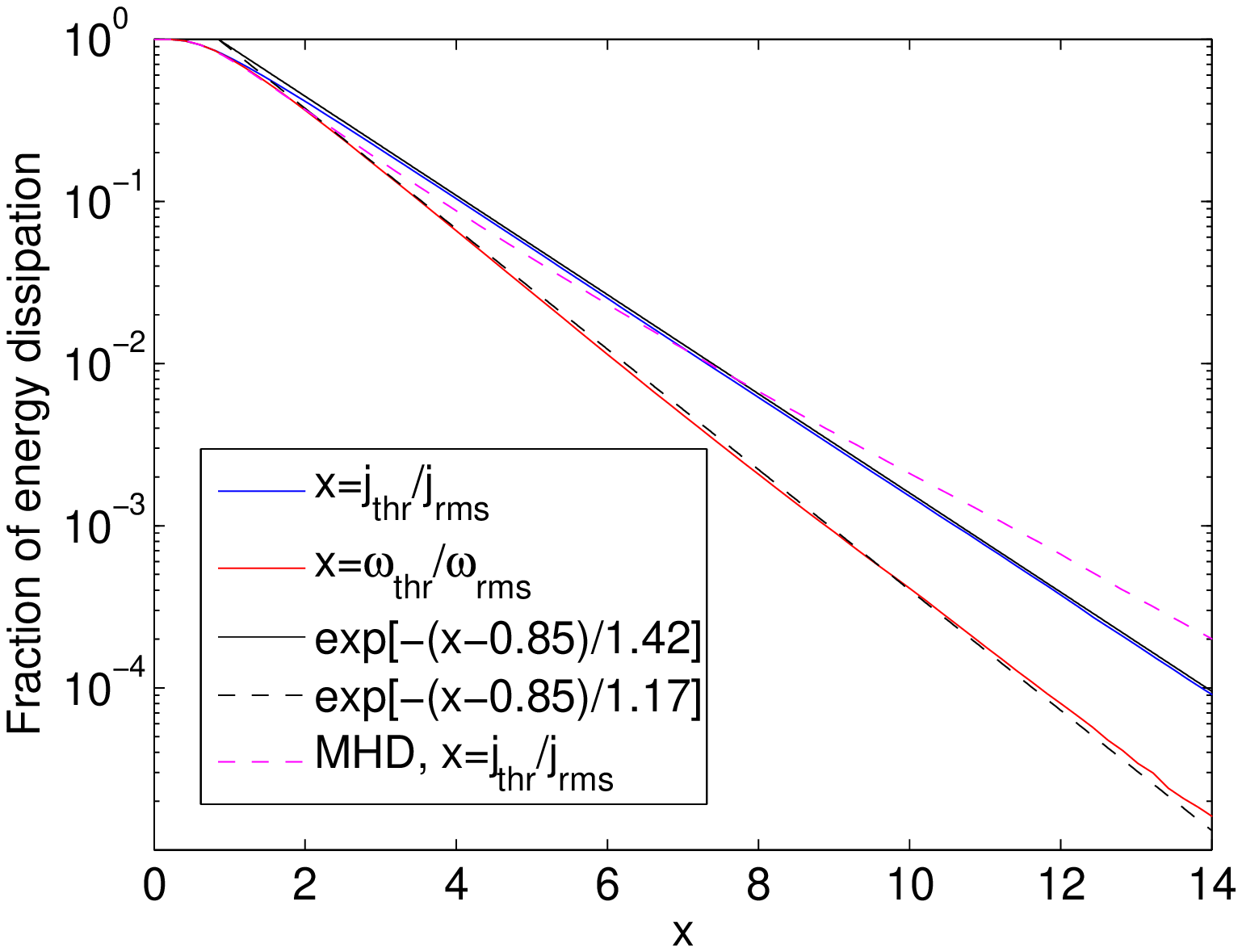}
 \centering
  \caption{\label{fig:dist_cum} Top panel: Fraction of total resistive dissipation (solid lines) and volume (dashed lines) in current density structures versus threshold $j_\text{thr}/j_\text{rms}$ for varying background fields ($B_0 = 0.03$ in blue, $B_0 = 0.1$ in red, and $B_0 = 0.005$ in green). Bottom panel: Exponential fits (black) to fraction of resistive dissipation in current density structures (blue) and fraction of viscous dissipation in vorticity structures (red) versus threshold. For comparison, the case for current density structures in standard MHD turbulence is also shown (dashed magenta line).}
 \end{figure}

We next describe the fraction of volume occupied by intermittent structures and their contribution to the overall energy dissipation, which is determined from cumulative distributions conditioned on thresholds in the current density and vorticity (denoted $j_\text{thr}$ and $\omega_\text{thr}$, respectively). In the first panel of Fig.~\ref{fig:dist_cum}, we show the resistive energy dissipation and volume filling fraction versus the normalized current density threshold $j_\text{thr}/j_\text{rms}$ (where $j_\text{rms}$ is the time-averaged rms current density). We find that the cumulative distributions are nearly identical for all three values of $B_0$. For large thresholds ($j_\text{thr}/j_\text{rms} \gg 1$), which is associated with intermittent current density structures, the fraction of energy dissipation declines exponentially, as $\exp{(-\lambda j_\text{thr}/j_\text{rms})}$ with $\lambda \approx 1/1.42$. We find that roughly $17\%$ of the resistive dissipation occurs in $1\%$ of the volume, and $50\%$ of the resistive dissipation in $7.5\%$ of the volume, indicating a high degree of inhomogeneity. In the second panel of Fig.~\ref{fig:dist_cum}, we show that the fraction of viscous dissipation (estimated by integrating $\nu \omega^2$) in vorticity structures at a threshold $\omega_\text{thr}/\omega_\text{rms}$ is also well fit by an exponential, with a slightly stronger decay constant, $\lambda \approx 1/1.17$. For comparison, we also show the resistive energy dissipation versus $j_\text{thr}/j_\text{rms}$ in simulations of standard homogeneous MHD turbulence with a background magnetic field that is comparable to turbulent fluctuations (taken from the data set in \cite{zhdankin_etal_2016}). We find that the two cases agree very well. In contrast, we note that standard MHD turbulence with a strong guide field is significantly more intermittent, with a smaller decay constant $\lambda \approx 1/3.3$ \citep[see, e.g.,][]{zhdankin_etal_2016b}.
 
 \begin{figure*}
\includegraphics[width=1.0\columnwidth]{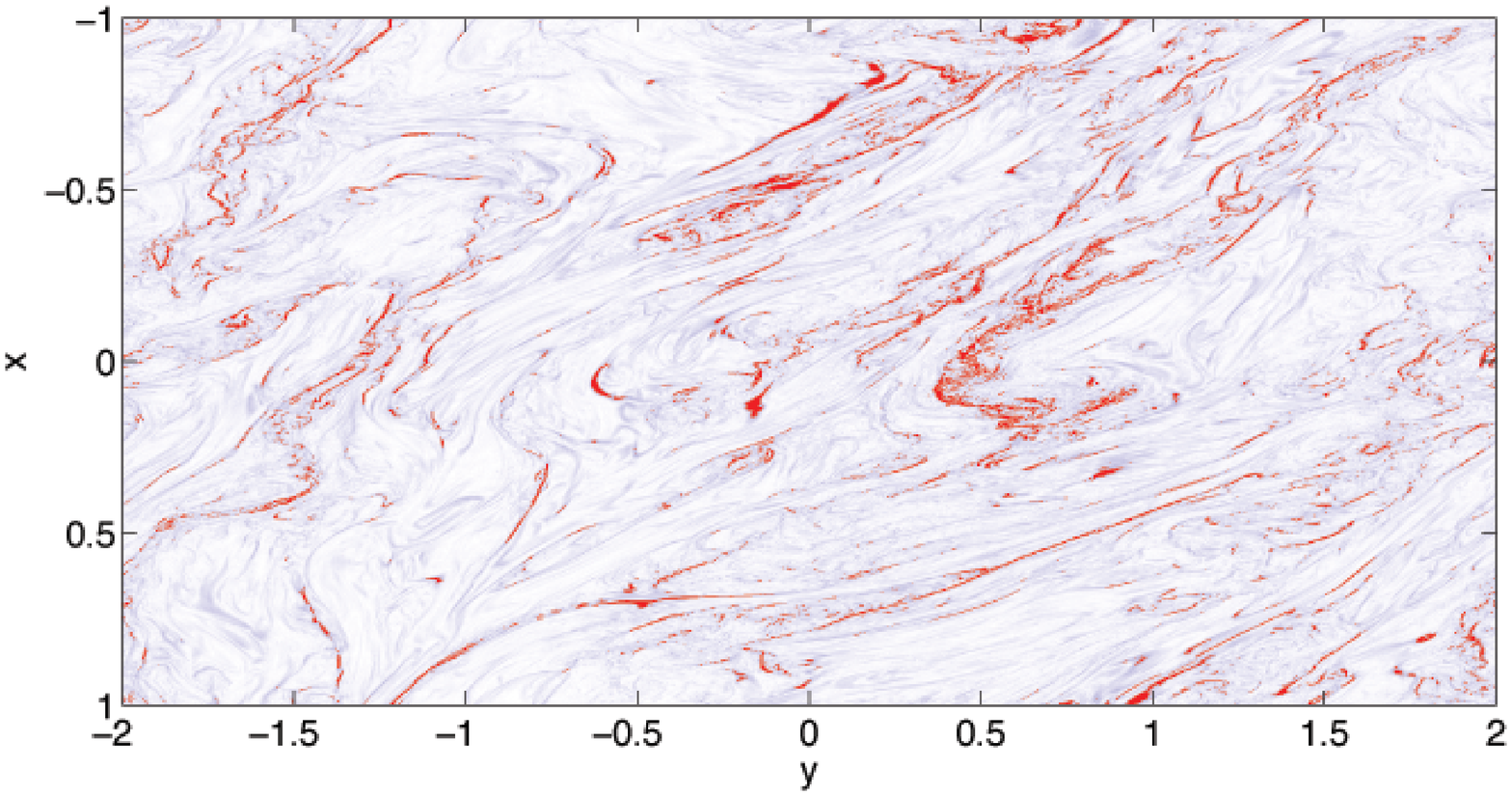}
\includegraphics[width=1.0\columnwidth]{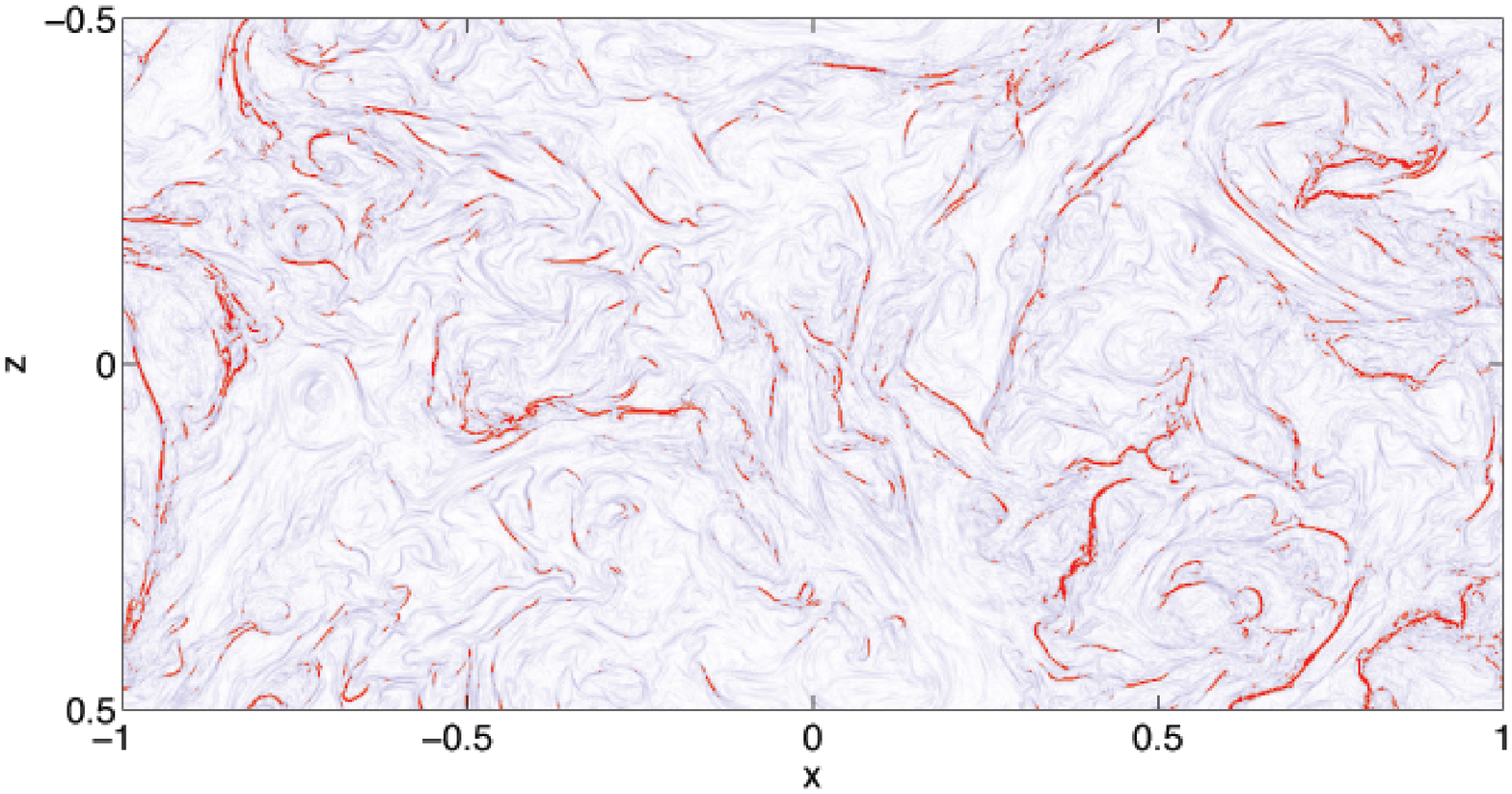}
 \centering
  \caption{\label{fig:images} Images of current density magnitude $j$ in $yx$ (azimuthal/radial) and $xz$ (radial/vertical) slices of the simulations. Points exceeding a threshold of $2.5 j_{\rm rms}$ are marked in red (note that a higher threshold is used in the analysis). Note the characteristic tilt of $\sim 17.5^\circ$ for structures in the $xy$ plane.}
 \end{figure*}
 
We now arrive at our statistical analysis of intermittent dissipative structures. We apply the methodology described in \cite{zhdankin_etal_2014} and mainly show the results for structures in the current density (i.e., current sheets). Each structure is defined to be a cluster of connected points with current density magnitudes exceeding a fixed threshold, $|\boldsymbol{j}| > j_\text{thr}$. For each current sheet, we measure the resistive energy dissipation rate ${\mathcal E} = \int dV \epsilon^\eta$ (integrated across the constituent points) and three characteristic scales: the length $L$, width $W$, and thickness $T$. The characteristic scales are measured in three orthogonal directions such that the length $L$ is the maximum distance between any two points in the structure, $W$ is the maximum distance between between any two points in the plane orthogonal to $\hat{\boldsymbol{L}}$ (the direction of $L$) coinciding with the point of maximum current density inside the structure, and $T$ is the distance across the structure in the direction $\hat{\boldsymbol{L}} \times \hat{\boldsymbol{W}}$ passing through the point of maximum current density inside the structure. The statistical results are insensitive to the threshold as long as $j_\text{thr} \gg j_\text{rms}$; we choose $j_\text{thr} = 4 j_\text{rms}$. We show the current density profile in representative $yx$ and $xz$ slices of the simulation in Fig.~\ref{fig:images}, with current sheets exceeding a threshold of $2.5 j_\text{rms}$ marked in red.
 
  \begin{figure*}
 \includegraphics[width=\columnwidth]{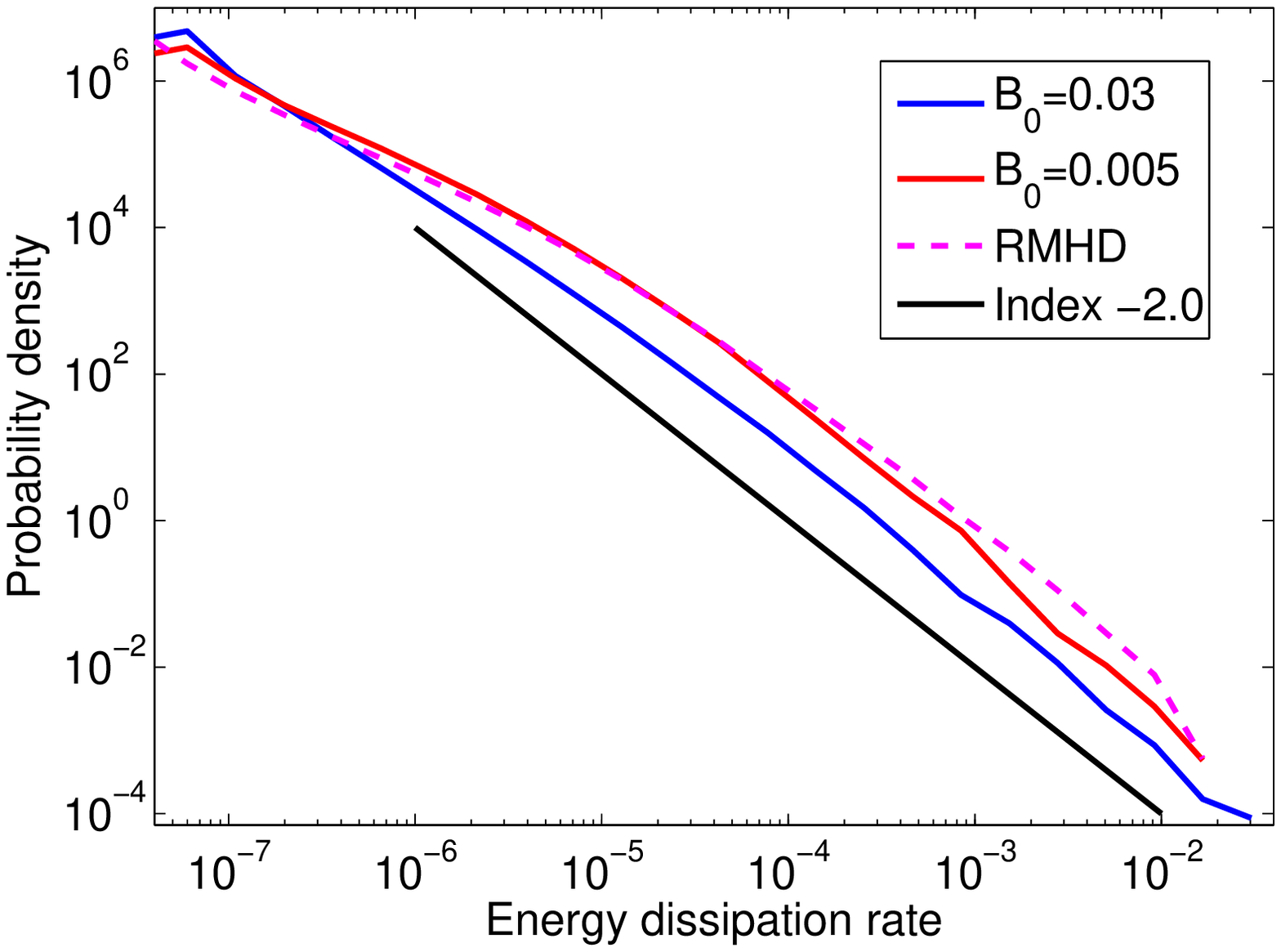}
\includegraphics[width=\columnwidth]{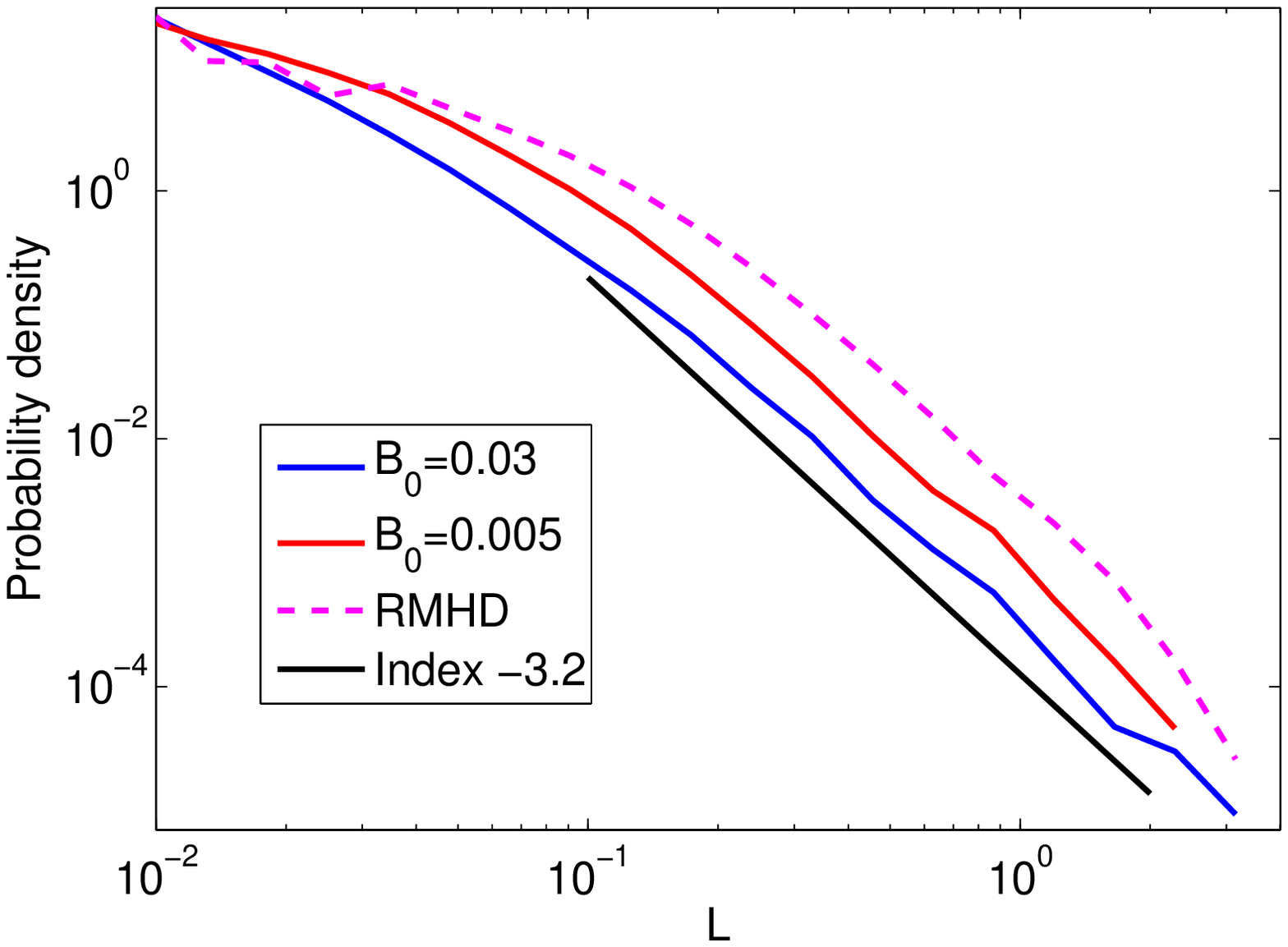}
\includegraphics[width=\columnwidth]{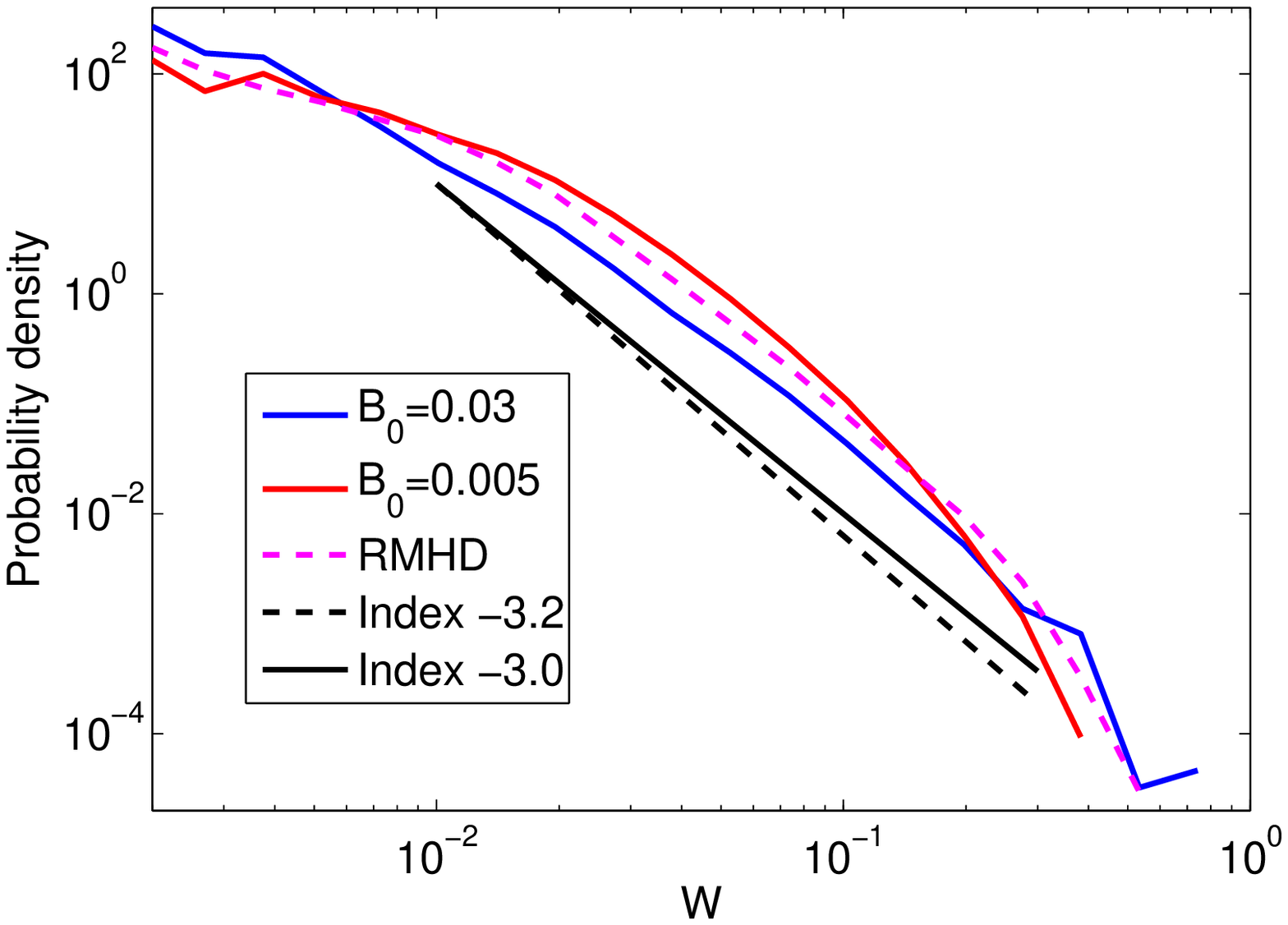}
\includegraphics[width=\columnwidth]{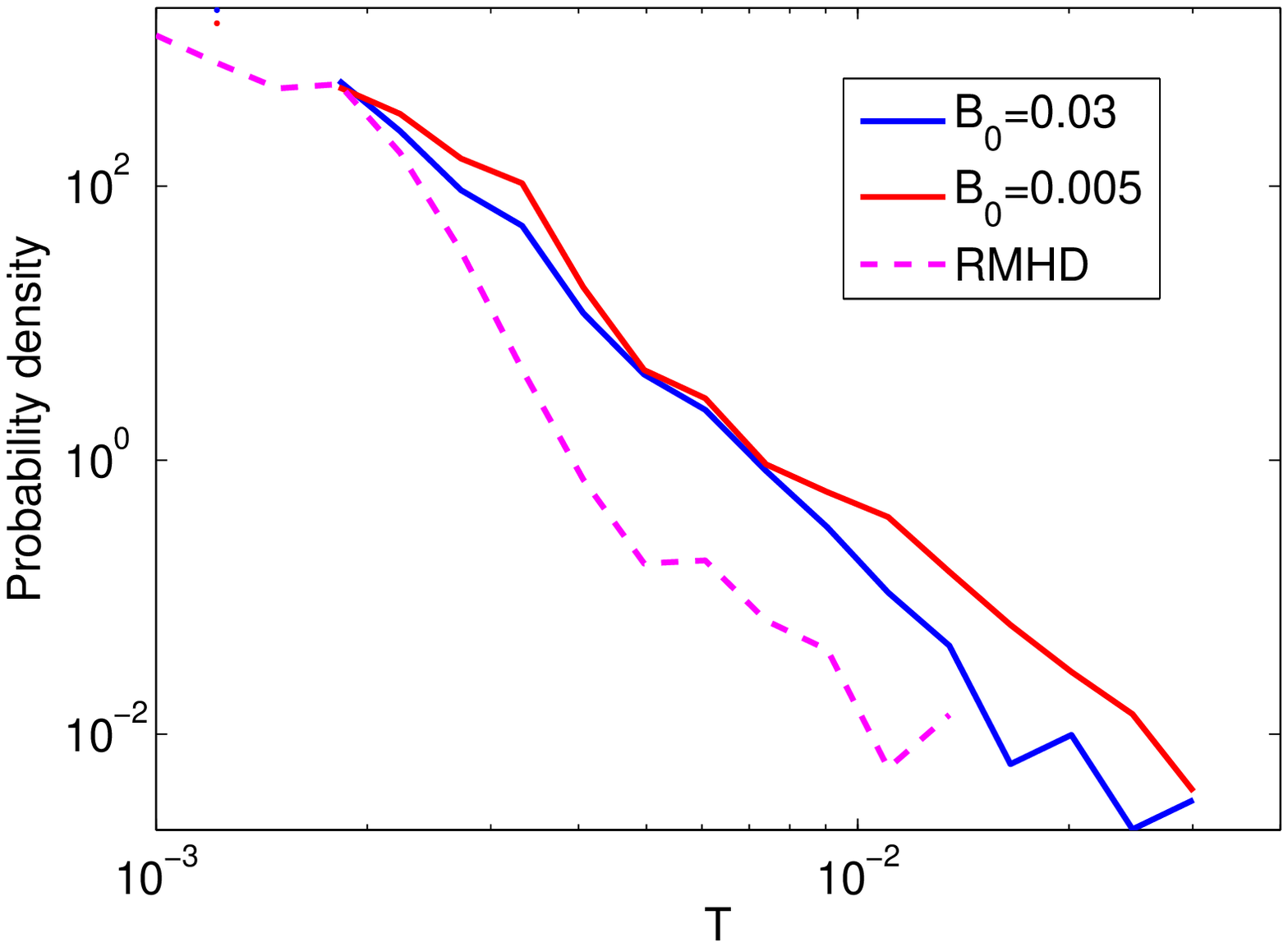} 
 \centering
  \caption{\label{fig:dists} Probability distributions for resistive energy dissipation rates ${\cal E}$, lengths $L$, widths $W$, and thicknesses $T$ of current sheets for $B_0 = 0.03$ (blue) and $B_0 = 0.005$ (red). For comparison, distributions from RMHD simulations are shown in magenta.}
 \end{figure*}
 
We show probability distributions for the current sheet measurements in Fig.~\ref{fig:dists}; for comparison, we also show distributions for the same measurements in a reduced MHD (RMHD) simulation with $Re = 3200$ and strong guide field $B_0 \approx 5 b_\text{rms}$ (taken from \cite{zhdankin_etal_2014, zhdankin_etal_2016b}). The distribution for energy dissipation rates ${\cal E}$ is consistent with a power law with a critical index of $-2.0$ across more than three decades, which implies that the populations of weak structures and strong structures both contribute equally to the overall energy dissipation. This critical index is also observed in the RMHD case. The distributions for lengths $L$ and widths $W$ also show power laws agreeing with RMHD, which has indices near $-3.2$ (the distribution for $W$ is possibly shallower, with index $-3.0$). Lengths and widths extend to scales comparable to the box size in the toroidal direction and the vertical direction, respectively. Thicknesses $T$ are peaked at small scales and do not exhibit a clear power law (although possibly a very steep one with index $\sim -5$). The distributions are very similar in all three simulations; in particular, the cases with $B_0 = 0.03$ and $B_0 = 0.01$ are nearly indistinguishable, while differences in the $B_0 = 0.005$ case (such as the shorter power law in the distribution for $W$) can be attributed to a shorter inertial range.

 \begin{figure}
\includegraphics[width=\columnwidth]{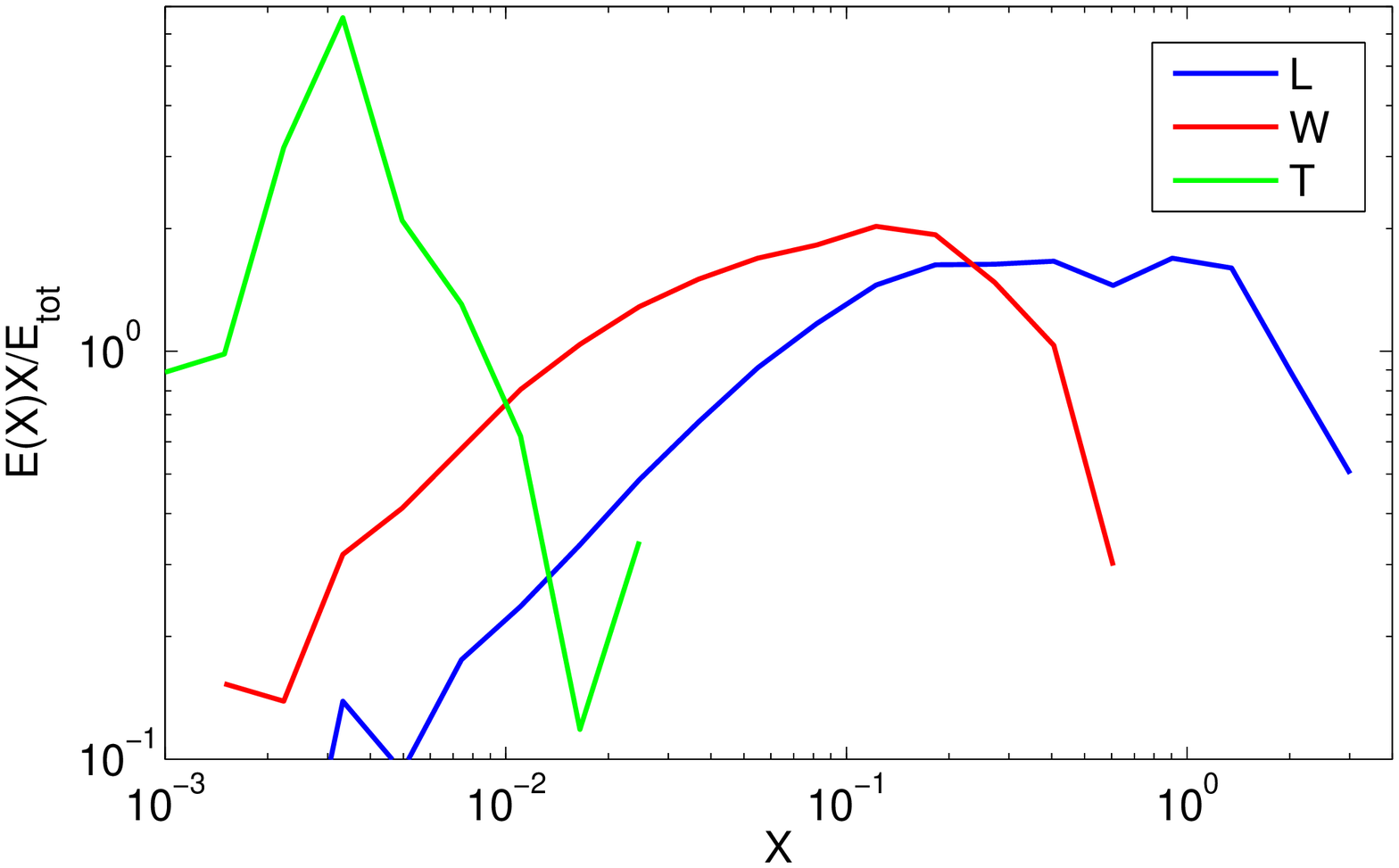}
 \centering
  \caption{\label{fig:dist_weighted} Compensated, dissipation-weighted distributions $E(X)X$ for current sheet scales $X \in \{ L , W, T \}$ (in blue, red, and green, respectively).}
 \end{figure}
 
One implication of the above distributions is that energy dissipation is spread across current sheets spanning a broad range of scales. This is evident when plotting compensated dissipation-weighted distributions $E(X)X$ for the current sheet sizes $X \in \{ L, W, T \}$, i.e., the distribution weighted by the energy dissipation rates and compensated by the given scale, which is shown in Fig.~\ref{fig:dist_weighted}. The broad peak of $E(L)L$ and $E(W)W$ reveal that energy dissipation is dominated by current sheets with lengths $0.1 L_z \lesssim L \lesssim 2 L_z$ and widths in the range $0.02 L_z \lesssim W \lesssim 0.3 L_z$. Meanwhile, thicknesses are peaked at small scales, $T \sim 3 \times 10^{-3} L_z$. We interpret these lengths and widths to be the inertial range in the corresponding directions, and the thickness to be the dissipation scale. We note that although the typical thicknesses are only marginally resolved in this case, a similar distribution was found in the much better resolved RMHD simulations \citep{zhdankin_etal_2016b}.
 
   \begin{figure}
\includegraphics[width=\columnwidth]{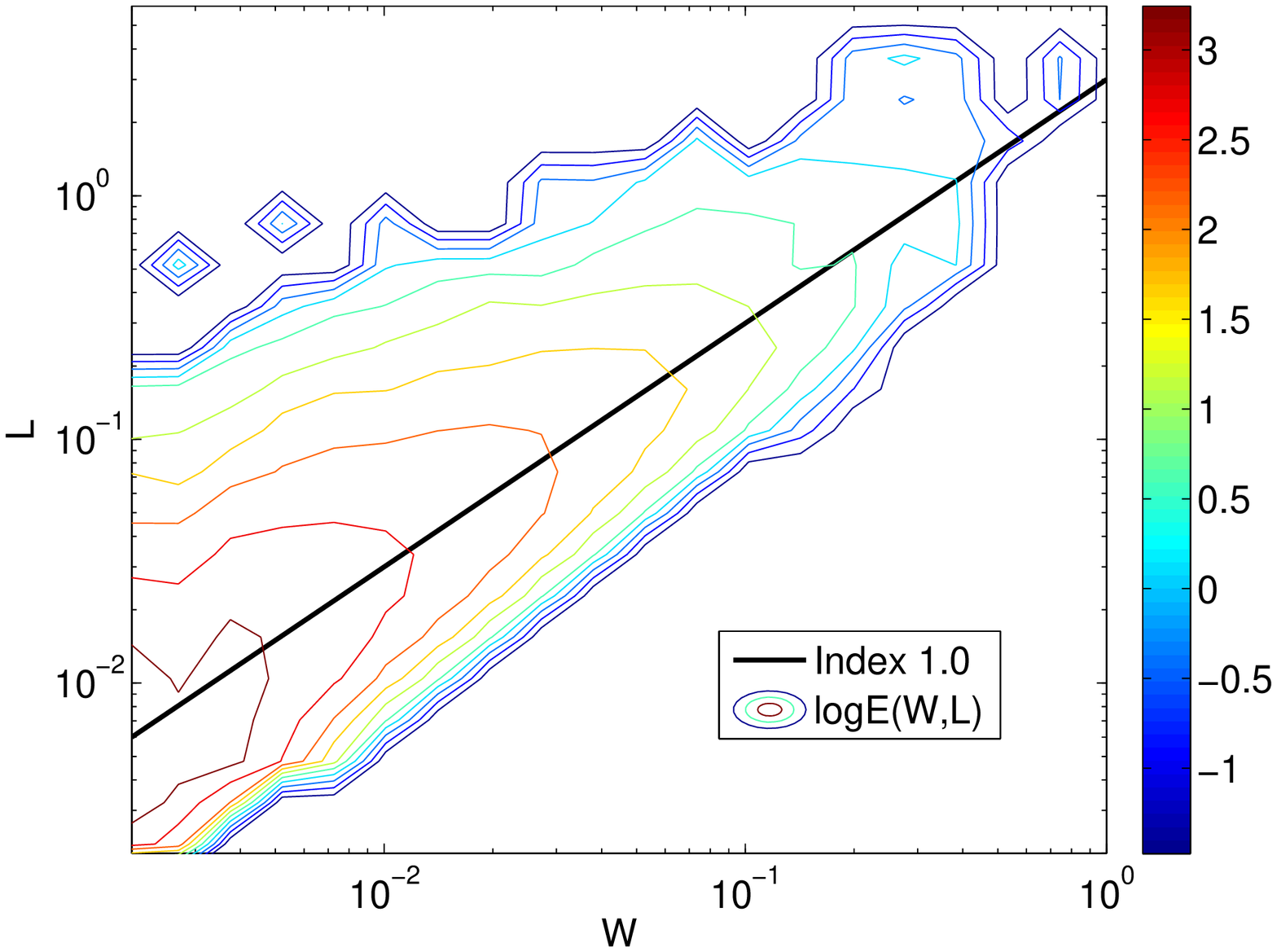}
\includegraphics[width=\columnwidth]{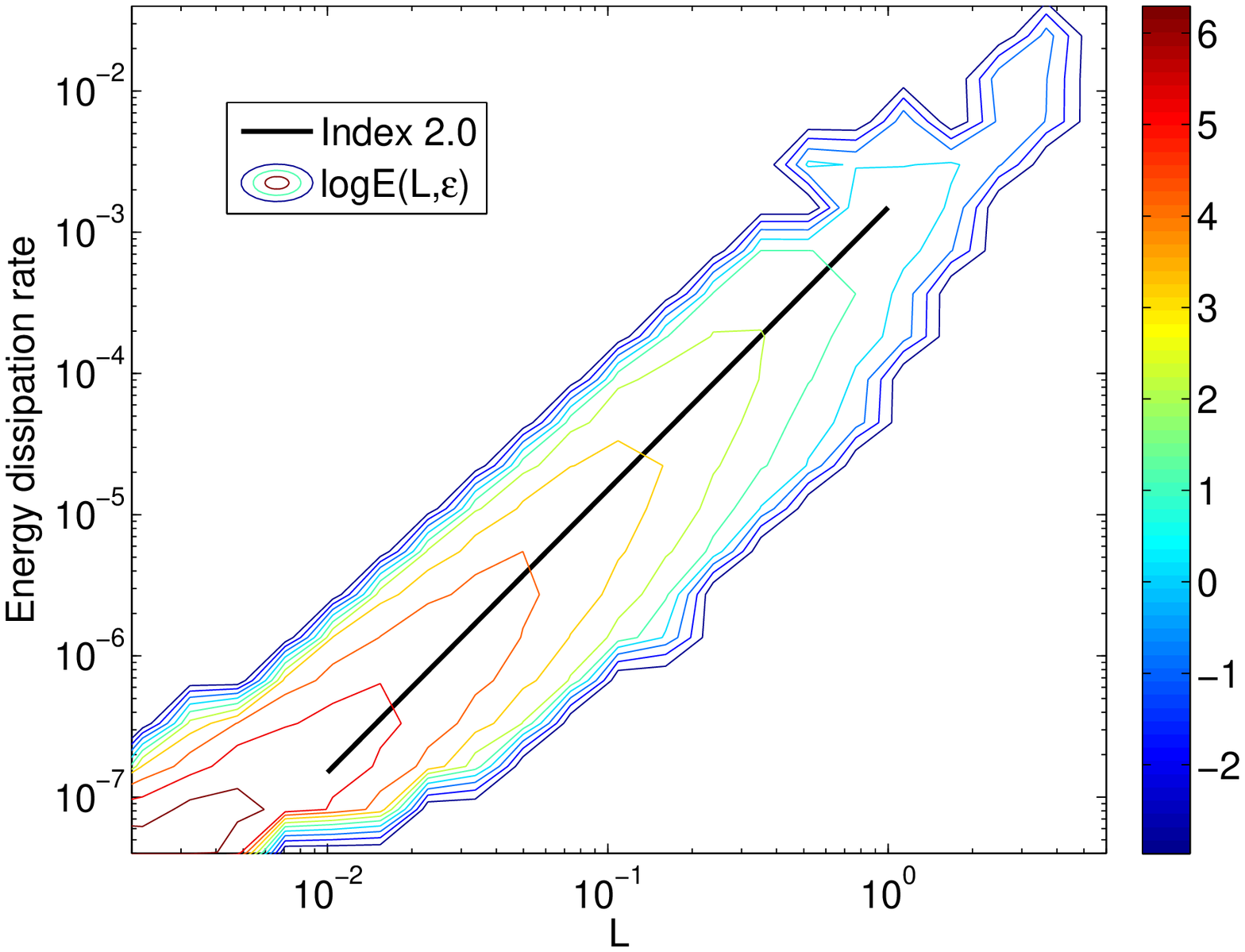}
 \centering
  \caption{\label{fig:dist_2d} Top panel: 2D dissipation-weighted distributions showing current sheet lengths $L$ versus widths $W$. Bottom panel: Similar for energy dissipation rates ${\cal E}$ versus $L$.}
 \end{figure}
 
The various measurements performed on the current sheets often exhibit correlations. These can be characterized from 2D distributions, as shown in Fig.~\ref{fig:dist_2d} for $L$ versus $W$ and for ${\cal E}$ versus $L$. For clarity, these 2D distributions are weighted by the dissipation rates. We find that $L \sim W$ and ${\cal E} \sim L^2$, similar to the scalings measured for RMHD \citep{zhdankin_etal_2016b}. The thickness varies over a much smaller range and does not exhibit a clear scaling.
 
   \begin{figure*}
   \includegraphics[width=2\columnwidth]{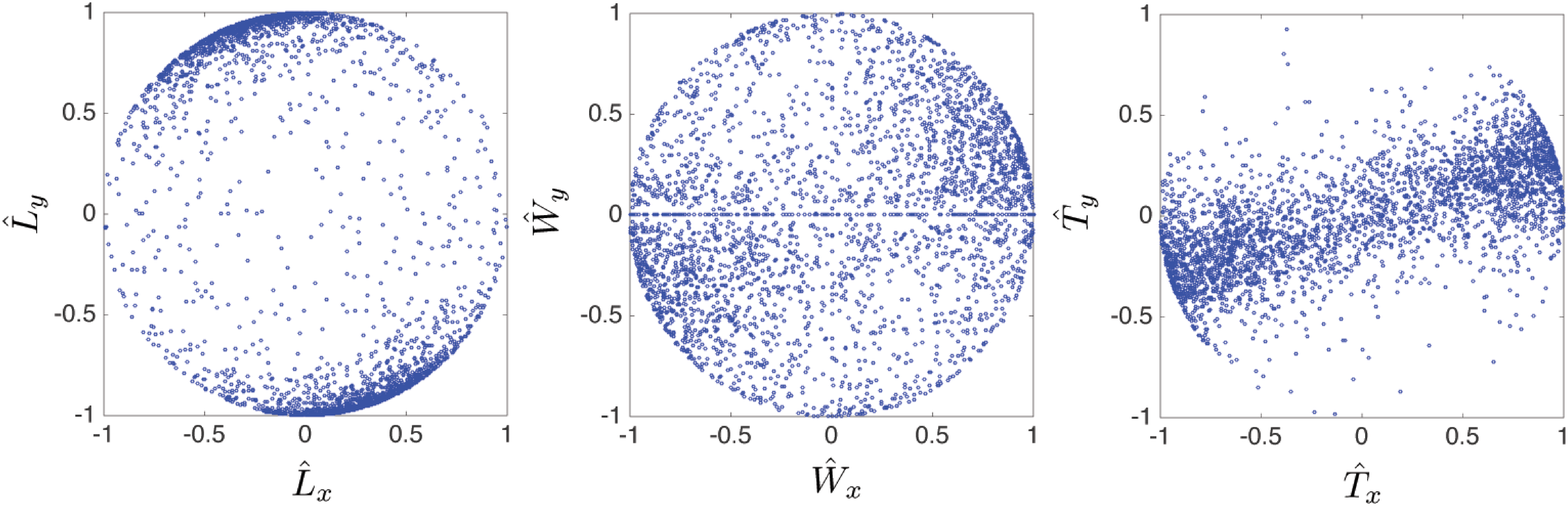}
 \centering
  \caption{\label{fig:orientation} Scatterplots of the components of $\hat{\boldsymbol{L}}$, $\hat{\boldsymbol{W}}$, and $\hat{\boldsymbol{T}}$ in the $xy$ plane (for intense current sheets with ${\cal E}/{\cal E}_{\rm tot} > 4 \times 10^{-5}$). A characteristic tilt of about $17.5^\circ$ counterclockwise is evident.}
 \end{figure*}
 
   \begin{figure*}
   \includegraphics[width=2\columnwidth]{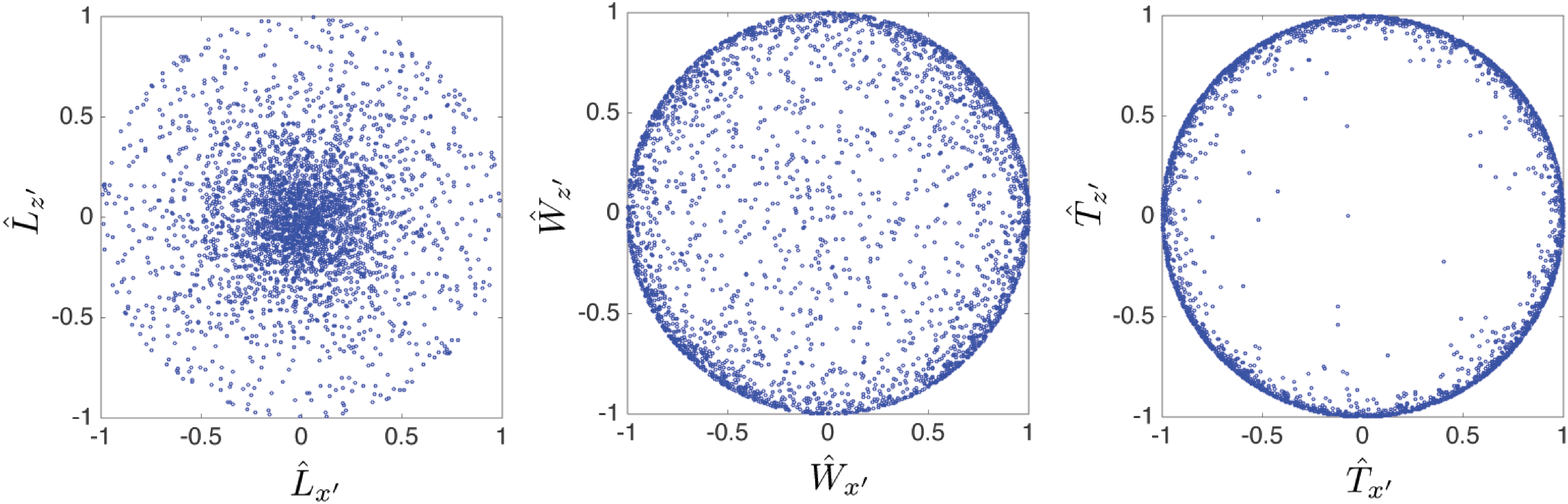}
 \centering
  \caption{\label{fig:orientation_rotated} Scatterplots of the components of $\hat{\boldsymbol{L}}$, $\hat{\boldsymbol{W}}$, and $\hat{\boldsymbol{T}}$ in the tilted $x'z'$ plane (for intense current sheets with ${\cal E}/{\cal E}_{\rm tot} > 4 \times 10^{-5}$). Isotropy around the (out-of-plane) $y'$ axis is evident.}
 \end{figure*}
 
Unlike standard MHD turbulence, where structures are elongated along the mean magnetic field, the typical orientation of structures in MRI-driven turbulence is not \it a priori \rm obvious. Therefore, we next characterize the orientation of current sheets by measuring direction vectors for the sizes: $\hat{\boldsymbol{L}}$, $\hat{\boldsymbol{W}}$, and $\hat{\boldsymbol{T}}$. We show scatterplots of the components of $\hat{\boldsymbol{L}}$, $\hat{\boldsymbol{W}}$, and $\hat{\boldsymbol{T}}$ in the $xy$ plane in Fig.~\ref{fig:orientation}; for clarity, only intense current sheets exceeding an energy dissipation rate ${\cal E}/{\cal E}_{\rm tot} > 4 \times 10^{-5}$ are shown. We find that the current sheets are typically extended in the azimuthal direction, but with a tilt of approximately $\theta_{\rm tilt} = 17.5^\circ$. To better characterize the orientations, we utilize the tilted coordinates $(x',y',z')$ obtained by rotating the original $(x,y,z)$ coordinate system by the angle $\theta_{\rm tilt}$ counterclockwise around the $z$ axis. We show the components of $\hat{\boldsymbol{L}}$, $\hat{\boldsymbol{W}}$, and $\hat{\boldsymbol{T}}$ in the tilted $x'z'$ plane in Fig.~\ref{fig:orientation_rotated}. Isotropy around the tilted $y'$ axis is evident from these scatterplots. Most of the current sheets have $\hat{L}_{y'} \approx \pm 1$, i.e., they are aligned with the tilt direction $\hat{\boldsymbol{y}}' \approx \hat{\boldsymbol{y}} \cos{\theta_{\rm tilt}} - \hat{\boldsymbol{x}} \sin{\theta_{\rm tilt}}$, while $\hat{\boldsymbol{W}}$ and $\hat{\boldsymbol{T}}$ are statistically isotropic in the plane perpendicular to $\hat{\boldsymbol{y}}'$. These orientations indicate that there may be an effective guide field in the $\hat{\boldsymbol{y}}'$ direction. This tilt angle is somewhat larger than the angle associated with the angular momentum transport coefficient noted in previous studies \citep[e.g.,][]{simon_etal_2012}. It however agrees well with distribution of the magnetic inclination angle, $\theta_{\rm inc} = \tan^{-1}{(B_x/B_y)}$, which we find to be strongly peaked near $\theta_{\rm inc} \sim - 18^\circ$ in our simulations.
 
  \begin{figure}
  \includegraphics[width=\columnwidth]{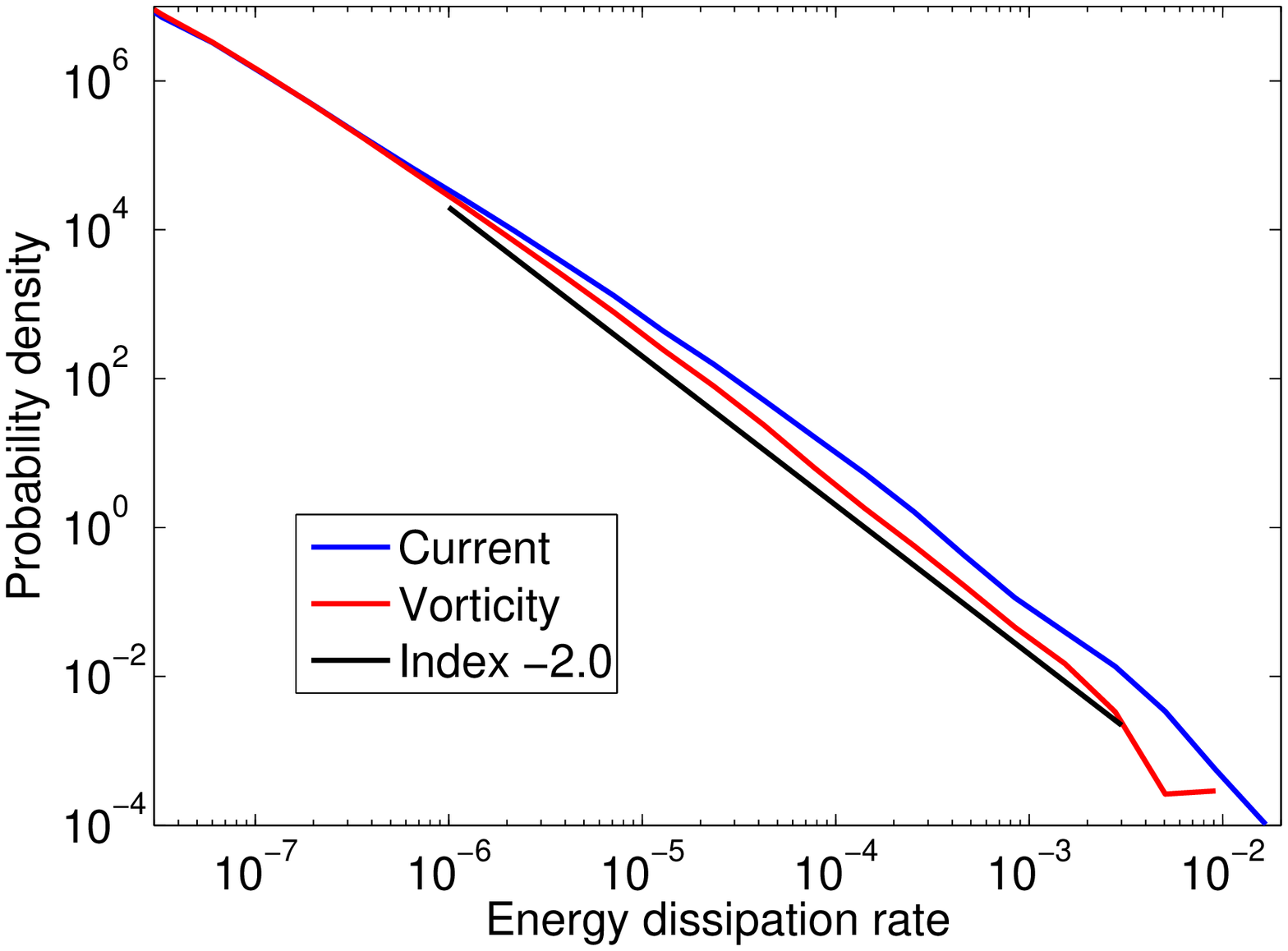}
 \centering
  \caption{\label{fig:vorticity} Distribution of viscous energy dissipation rates for vorticity sheets compared to resistive energy dissipation rates for current sheets.}
 \end{figure}
 
In addition to the above analysis of current sheets, we performed a similar analysis for vorticity sheets by identifying structures that exceed a vorticity threshold, $|\boldsymbol{\omega}| > \omega_\text{thr}$. The statistical results are essentially the same as for the current sheets, in agreement with the standard case \citep{zhdankin_etal_2016b}. For example, we compare the distribution of viscous energy dissipation rates (estimated by integrating $\nu \omega^2$ across the structure) for vorticity structures to resistive dissipation rates of current sheets in Fig.~\ref{fig:vorticity}. Vorticity structures tend to be somewhat weaker and smaller (with distributions for length and width being shifted to smaller scales by less than a factor of 2), but otherwise have quantitatively similar distributions. The orientations of the vorticity structures are also very similar to current sheets, being aligned with the same tilt axis.

\section{Conclusions}

In this work, we investigated the small-scale structure and intermittency of energy dissipation in a set of MHD simulations of MRI-driven turbulence. We found that MRI-driven turbulence leaves an imprint on small-scale quantities (e.g., current density, vorticity, dissipation rates) that is nearly indistinguishable from simulations of standard homogeneous MHD turbulence with a uniform guide field. This is in contrast to previous studies which focused on inertial-range quantities such as the energy spectrum and found little resemblence to standard turbulence \citep{fromang_2010, lesur_longaretti_2011}, likely due to the effects of large-scale shear and broadband energy injection from the MRI in a relatively small domain. Our results indicate that the small-scale structure is insensitive to these effects, and can be understood as standard MHD turbulence with a self-consistently generated guide field that is slightly tilted from the toroidal direction. Surprisingly, even the inertial-range lengths and widths of intermittent structures agree with the standard case. Other quantities such as the energy spectra may duplicate the results from standard MHD turbulence once simulations become sufficiently large.

If the above picture is correct, then the small-scale turbulence in accretion disks can be modeled with standard MHD turbulence phenomenology \citep{goldreich_sridhar_1995, boldyrev_2005}. One implication is that the dynamics at kinetic scales, responsible for plasma heating and particle acceleration, will be insensitive to the large-scale accretion disk physics. For practical applications, future work is required to verify that these conclusions are insensitive to stratification, radiative effects, and global geometry. It also remains to be shown that the results are unaffected by compressive fluctuations \citep{federrath_etal_2011, federrath_etal_2014}, which may form shocks as an alternative route to intermittent dissipation \citep{schmidt_etal_2008}. For instance, it is known that the excitation of mildy subsonic MHD turbulence in Keplerian shear flows excites non-axisymmetric spiral density waves which are ultimately dissipated as shocks \citep{heinemann_papaloizou_2009a, heinemann_papaloizou_2009b}.

In the present study, we used numerical simulations corresponding to magnetic Prandtl number $Pm=1$. This was done to maximize the inertial range of turbulence for the available numerical resolution. In our future work we will address the cases of $Pm<1$ and $Pm >1$. In the latter case, MRI-driven turbulence may be excited solely by the nonlinear magnetic dynamo action, that is, without an imposed magnetic flux \citep[e.g.,][]{lesur_longaretti_2007}. Whether this affects the properties of MRI-driven turbulence is an interesting question.

\section*{Acknowledgements}

We thank Dmitri Uzdensky and Mitch Begelman for fruitful discussions and for comments on the manuscript. VZ acknowledges support from NSF grant AST-1411879. JW is supported by the Wisconsin Alumni Research Foundation at the University of Wisconsin-Madison. SB is partly supported by NSF grant AGS-1261659, and by the Vilas Associates Award from the University of Wisconsin-Madison. Simulations were performed at the Texas Advanced Computing Center (TACC) at the University of Texas at Austin under the NSF-Teragrid Project TG-PHY110016.






\begin{thebibliography}{}
\makeatletter
\relax
\def\mn@urlcharsother{\let\do\@makeother \do\$\do\&\do\#\do\^\do\_\do\%\do\~}
\def\mn@doi{\begingroup\mn@urlcharsother \@ifnextchar [ {\mn@doi@}
  {\mn@doi@[]}}
\def\mn@doi@[#1]#2{\def\@tempa{#1}\ifx\@tempa\@empty \href
  {http://dx.doi.org/#2} {doi:#2}\else \href {http://dx.doi.org/#2} {#1}\fi
  \endgroup}
\def\mn@eprint#1#2{\mn@eprint@#1:#2::\@nil}
\def\mn@eprint@arXiv#1{\href {http://arxiv.org/abs/#1} {{\tt arXiv:#1}}}
\def\mn@eprint@dblp#1{\href {http://dblp.uni-trier.de/rec/bibtex/#1.xml}
  {dblp:#1}}
\def\mn@eprint@#1:#2:#3:#4\@nil{\def\@tempa {#1}\def\@tempb {#2}\def\@tempc
  {#3}\ifx \@tempc \@empty \let \@tempc \@tempb \let \@tempb \@tempa \fi \ifx
  \@tempb \@empty \def\@tempb {arXiv}\fi \@ifundefined
  {mn@eprint@\@tempb}{\@tempb:\@tempc}{\expandafter \expandafter \csname
  mn@eprint@\@tempb\endcsname \expandafter{\@tempc}}}

\bibitem[\protect\citeauthoryear{Baganoff et~al.,}{Baganoff
  et~al.}{2001}]{baganoff_etal_2001}
Baganoff F.,  et~al., 2001, Nature, 413, 45

\bibitem[\protect\citeauthoryear{Bai \& Stone}{Bai \&
  Stone}{2013}]{bai_stone_2013}
Bai X.-N.,  Stone J.~M.,  2013, The Astrophysical Journal, 767, 30

\bibitem[\protect\citeauthoryear{Balbus \& Hawley}{Balbus \&
  Hawley}{1991}]{balbus_hawley_1991}
Balbus S.~A.,  Hawley J.~F.,  1991, The Astrophysical Journal, 376, 214

\bibitem[\protect\citeauthoryear{Balbus \& Hawley}{Balbus \&
  Hawley}{1998}]{balbus_hawley_1998}
Balbus S.~A.,  Hawley J.~F.,  1998, Reviews of modern physics, 70, 1

\bibitem[\protect\citeauthoryear{Biskamp}{Biskamp}{1995}]{biskamp_1995}
Biskamp D.,  1995, Chaos, Solitons \& Fractals, 5, 1779

\bibitem[\protect\citeauthoryear{Biskamp}{Biskamp}{2003}]{biskamp2003}
Biskamp D.,  2003, Magnetohydrodynamic turbulence.
Cambridge Univ Pr

\bibitem[\protect\citeauthoryear{Bodo, Mignone, Cattaneo, Rossi  \&
  Ferrari}{Bodo et~al.}{2008}]{bodo_etal_2008}
Bodo G.,  Mignone A.,  Cattaneo F.,  Rossi P.,   Ferrari A.,  2008, Astronomy
  \& Astrophysics, 487, 1

\bibitem[\protect\citeauthoryear{Bodo, Cattaneo, Ferrari, Mignone  \&
  Rossi}{Bodo et~al.}{2011}]{bodo_etal_2011}
Bodo G.,  Cattaneo F.,  Ferrari A.,  Mignone A.,   Rossi P.,  2011, The
  Astrophysical Journal, 739, 82

\bibitem[\protect\citeauthoryear{Boldyrev}{Boldyrev}{2005}]{boldyrev_2005}
Boldyrev S.,  2005, The Astrophysical Journal Letters, 626, L37

\bibitem[\protect\citeauthoryear{Boldyrev}{Boldyrev}{2006}]{boldyrev2006}
Boldyrev S.,  2006, Physical Review Letters, 96, 115002

\bibitem[\protect\citeauthoryear{Brandenburg, Nordlund, Stein  \&
  Torkelsson}{Brandenburg et~al.}{1995}]{brandenburg_etal_1995}
Brandenburg A.,  Nordlund A.,  Stein R.~F.,   Torkelsson U.,  1995, The
  Astrophysical Journal, 446, 741

\bibitem[\protect\citeauthoryear{Chandran, Schekochihin  \& Mallet}{Chandran
  et~al.}{2015}]{chandran_etal_2015}
Chandran B.~D.,  Schekochihin A.~A.,   Mallet A.,  2015, The Astrophysical
  Journal, 807, 39

\bibitem[\protect\citeauthoryear{Federrath, Chabrier, Schober, Banerjee,
  Klessen  \& Schleicher}{Federrath et~al.}{2011}]{federrath_etal_2011}
Federrath C.,  Chabrier G.,  Schober J.,  Banerjee R.,  Klessen R.~S.,
  Schleicher D.~R.,  2011, Physical Review Letters, 107, 114504

\bibitem[\protect\citeauthoryear{Federrath, Schober, Bovino  \&
  Schleicher}{Federrath et~al.}{2014}]{federrath_etal_2014}
Federrath C.,  Schober J.,  Bovino S.,   Schleicher D.~R.,  2014, The
  Astrophysical Journal Letters, 797, L19

\bibitem[\protect\citeauthoryear{Fleming, Stone  \& Hawley}{Fleming
  et~al.}{2000}]{fleming_etal_2000}
Fleming T.~P.,  Stone J.~M.,   Hawley J.~F.,  2000, The Astrophysical Journal,
  530, 464

\bibitem[\protect\citeauthoryear{Frisch}{Frisch}{1995}]{frisch1995}
Frisch U.,  1995, Turbulence: The Legacy of AN Kolmogorov.
Cambridge Univ. Press

\bibitem[\protect\citeauthoryear{Fromang}{Fromang}{2010}]{fromang_2010}
Fromang S.,  2010, Astronomy \& Astrophysics, 514, L5

\bibitem[\protect\citeauthoryear{Fromang, Papaloizou, Lesur  \&
  Heinemann}{Fromang et~al.}{2007}]{fromang_etal_2007}
Fromang S.,  Papaloizou J.,  Lesur G.,   Heinemann T.,  2007, Astronomy \&
  Astrophysics, 476, 1123

\bibitem[\protect\citeauthoryear{Goldreich \& Sridhar}{Goldreich \&
  Sridhar}{1995}]{goldreich_sridhar_1995}
Goldreich P.,  Sridhar S.,  1995, The Astrophysical Journal, 438, 763

\bibitem[\protect\citeauthoryear{Heinemann \& Papaloizou}{Heinemann \&
  Papaloizou}{2009a}]{heinemann_papaloizou_2009a}
Heinemann T.,  Papaloizou J.,  2009a, Monthly Notices of the Royal Astronomical
  Society, 397, 52

\bibitem[\protect\citeauthoryear{Heinemann \& Papaloizou}{Heinemann \&
  Papaloizou}{2009b}]{heinemann_papaloizou_2009b}
Heinemann T.,  Papaloizou J.,  2009b, Monthly Notices of the Royal Astronomical
  Society, 397, 64

\bibitem[\protect\citeauthoryear{Kolmogorov}{Kolmogorov}{1962}]{kolmogorov_1962}
Kolmogorov A.~N.,  1962, Journal of Fluid Mechanics, 13, 82

\bibitem[\protect\citeauthoryear{{Kunz}, {Stone}  \& {Quataert}}{{Kunz}
  et~al.}{2016}]{kunz2016}
{Kunz} M.~W.,  {Stone} J.~M.,   {Quataert} E.,  2016, \mn@doi [Physical Review
  Letters] {10.1103/PhysRevLett.117.235101}, \href
  {http://adsabs.harvard.edu/abs/2016PhRvL.117w5101K} {117, 235101}

\bibitem[\protect\citeauthoryear{Lesur \& Longaretti}{Lesur \&
  Longaretti}{2007}]{lesur_longaretti_2007}
Lesur G.,  Longaretti P.-Y.,  2007, Monthly Notices of the Royal Astronomical
  Society, 378, 1471

\bibitem[\protect\citeauthoryear{Lesur \& Longaretti}{Lesur \&
  Longaretti}{2011}]{lesur_longaretti_2011}
Lesur G.,  Longaretti P.-Y.,  2011, Astronomy \& Astrophysics, 528, A17

\bibitem[\protect\citeauthoryear{Mallet \& Schekochihin}{Mallet \&
  Schekochihin}{2016}]{mallet_etal_2016}
Mallet A.,  Schekochihin A.,  2016, Monthly Notices of the Royal Astronomical
  Society, p. stw3251

\bibitem[\protect\citeauthoryear{Markoff, Falcke, Yuan  \& Biermann}{Markoff
  et~al.}{2001}]{markoff_etal_2001}
Markoff S.,  Falcke H.,  Yuan F.,   Biermann P.~L.,  2001, Astronomy \&
  Astrophysics, 379, L13

\bibitem[\protect\citeauthoryear{{Mason}, {Perez}, {Boldyrev}  \&
  {Cattaneo}}{{Mason} et~al.}{2012}]{mason2012}
{Mason} J.,  {Perez} J.~C.,  {Boldyrev} S.,   {Cattaneo} F.,  2012, \mn@doi
  [Physics of Plasmas] {10.1063/1.3694123}, \href
  {http://adsabs.harvard.edu/abs/2012PhPl...19e5902M} {19, 055902}

\bibitem[\protect\citeauthoryear{Matthaeus, Wan, Servidio, Greco, Osman,
  Oughton  \& Dmitruk}{Matthaeus et~al.}{2015}]{matthaeus_etal_2015}
Matthaeus W.,  Wan M.,  Servidio S.,  Greco A.,  Osman K.,  Oughton S.,
  Dmitruk P.,  2015, Philosophical Transactions of the Royal Society of London
  A: Mathematical, Physical and Engineering Sciences, 373, 20140154

\bibitem[\protect\citeauthoryear{{McClintock} \& {Remillard}}{{McClintock} \&
  {Remillard}}{2006}]{mcclintock_remillard_2006}
{McClintock} J.~E.,  {Remillard} R.~A.,  2006, {Black hole binaries}.
pp 157--213

\bibitem[\protect\citeauthoryear{{McNally}, {Hubbard}, {Yang}  \& {Mac
  Low}}{{McNally} et~al.}{2014}]{mcnally_etal_2014}
{McNally} C.~P.,  {Hubbard} A.,  {Yang} C.-C.,   {Mac Low} M.-M.,  2014,
  \mn@doi [The Astrophysical Journal] {10.1088/0004-637X/791/1/62}, 791, 62

\bibitem[\protect\citeauthoryear{Merrifield, M{\"u}ller, Chapman  \&
  Dendy}{Merrifield et~al.}{2005}]{merrifield_etal_2005}
Merrifield J.~A.,  M{\"u}ller W.-C.,  Chapman S.~C.,   Dendy R.~O.,  2005,
  Physics of Plasmas, 12, 022301

\bibitem[\protect\citeauthoryear{Murphy \& Pessah}{Murphy \&
  Pessah}{2015}]{murphy_pessah_2015}
Murphy G.~C.,  Pessah M.~E.,  2015, The Astrophysical Journal, 802, 139

\bibitem[\protect\citeauthoryear{Nauman \& Blackman}{Nauman \&
  Blackman}{2014}]{nauman_blackman_2014}
Nauman F.,  Blackman E.~G.,  2014, Monthly Notices of the Royal Astronomical
  Society, 441, 1855

\bibitem[\protect\citeauthoryear{{Perez}, {Mason}, {Boldyrev}  \&
  {Cattaneo}}{{Perez} et~al.}{2012}]{perez2012}
{Perez} J.~C.,  {Mason} J.,  {Boldyrev} S.,   {Cattaneo} F.,  2012, \mn@doi
  [Physical Review X] {10.1103/PhysRevX.2.041005}, \href
  {http://adsabs.harvard.edu/abs/2012PhRvX...2d1005P} {2, 041005}

\bibitem[\protect\citeauthoryear{{Salvesen}, {Simon}, {Armitage}  \&
  {Begelman}}{{Salvesen} et~al.}{2016}]{salvesen_etal_2016}
{Salvesen} G.,  {Simon} J.~B.,  {Armitage} P.~J.,   {Begelman} M.~C.,  2016,
  Monthly Notices of the Royal Astronomical Society, 457, 857

\bibitem[\protect\citeauthoryear{Schmidt, Federrath  \& Klessen}{Schmidt
  et~al.}{2008}]{schmidt_etal_2008}
Schmidt W.,  Federrath C.,   Klessen R.,  2008, Physical Review Letters, 101,
  194505

\bibitem[\protect\citeauthoryear{{Shi}, {Stone}  \& {Huang}}{{Shi}
  et~al.}{2016}]{shi_etal_2016}
{Shi} J.-M.,  {Stone} J.~M.,   {Huang} C.~X.,  2016, Monthly Notices of the
  Royal Astronomical Society, 456, 2273

\bibitem[\protect\citeauthoryear{Simon, Hawley  \& Beckwith}{Simon
  et~al.}{2011}]{simon_etal_2011}
Simon J.~B.,  Hawley J.~F.,   Beckwith K.,  2011, The Astrophysical Journal,
  730, 94

\bibitem[\protect\citeauthoryear{Simon, Beckwith  \& Armitage}{Simon
  et~al.}{2012}]{simon_etal_2012}
Simon J.~B.,  Beckwith K.,   Armitage P.~J.,  2012, Monthly Notices of the
  Royal Astronomical Society, 422, 2685

\bibitem[\protect\citeauthoryear{Sorathia, Reynolds, Stone  \&
  Beckwith}{Sorathia et~al.}{2012}]{sorathia_etal_2012}
Sorathia K.~A.,  Reynolds C.~S.,  Stone J.~M.,   Beckwith K.,  2012, The
  Astrophysical Journal, 749, 189

\bibitem[\protect\citeauthoryear{Sreenivasan \& Kailasnath}{Sreenivasan \&
  Kailasnath}{1993}]{sreenivasan_1993}
Sreenivasan K.,  Kailasnath P.,  1993, Physics of Fluids A: Fluid Dynamics
  (1989-1993), 5, 512

\bibitem[\protect\citeauthoryear{{Walker}, {Lesur}  \& {Boldyrev}}{{Walker}
  et~al.}{2016}]{walker_etal_2016}
{Walker} J.,  {Lesur} G.,   {Boldyrev} S.,  2016, \mn@doi [Monthly Notices of
  the Royal Astronomical Society: Letters] {10.1093/mnrasl/slv200}, 457, L39

\bibitem[\protect\citeauthoryear{Wan, Matthaeus, Roytershteyn, Parashar, Wu  \&
  Karimabadi}{Wan et~al.}{2016}]{wan_etal_2016}
Wan M.,  Matthaeus W.,  Roytershteyn V.,  Parashar T.,  Wu P.,   Karimabadi H.,
   2016, Physics of Plasmas (1994-present), 23, 042307

\bibitem[\protect\citeauthoryear{{Zhdankin}, {Uzdensky}, {Perez}  \&
  {Boldyrev}}{{Zhdankin} et~al.}{2013}]{zhdankin_etal_2013}
{Zhdankin} V.,  {Uzdensky} D.~A.,  {Perez} J.~C.,   {Boldyrev} S.,  2013, The
  Astrophysical Journal, 771, 124

\bibitem[\protect\citeauthoryear{{Zhdankin}, {Boldyrev}, {Perez}  \&
  {Tobias}}{{Zhdankin} et~al.}{2014}]{zhdankin_etal_2014}
{Zhdankin} V.,  {Boldyrev} S.,  {Perez} J.~C.,   {Tobias} S.~M.,  2014, The
  Astrophysical Journal, 795, 127

\bibitem[\protect\citeauthoryear{Zhdankin, Uzdensky  \& Boldyrev}{Zhdankin
  et~al.}{2015a}]{zhdankin_etal_2015a}
Zhdankin V.,  Uzdensky D.~A.,   Boldyrev S.,  2015a, Physical review letters,
  114, 065002

\bibitem[\protect\citeauthoryear{Zhdankin, Uzdensky  \& Boldyrev}{Zhdankin
  et~al.}{2015b}]{zhdankin_etal_2015b}
Zhdankin V.,  Uzdensky D.~A.,   Boldyrev S.,  2015b, The Astrophysical Journal,
  811, 6

\bibitem[\protect\citeauthoryear{Zhdankin, Boldyrev  \& Uzdensky}{Zhdankin
  et~al.}{2016a}]{zhdankin_etal_2016b}
Zhdankin V.,  Boldyrev S.,   Uzdensky D.~A.,  2016a, Physics of Plasmas
  (1994-present), 23, 055705

\bibitem[\protect\citeauthoryear{{Zhdankin}, {Boldyrev}  \& {Chen}}{{Zhdankin}
  et~al.}{2016b}]{zhdankin_etal_2016}
{Zhdankin} V.,  {Boldyrev} S.,   {Chen} C.~H.~K.,  2016b, \mn@doi [Monthly
  Notices of the Royal Astronomical Society: Letters] {10.1093/mnrasl/slv208},
  457, L69

\makeatother
\end{thebibliography}


\bsp	
\label{lastpage}
\end{document}